%% file: main.tex
\documentclass[11pt]{article}

\usepackage[english]{babel}   
\usepackage{setspace}         
\usepackage[a4paper,top=2cm,bottom=2cm,
            left=2cm,right=2cm,marginparwidth=1.75cm]{geometry}

\usepackage{amsmath,amssymb,amsthm,amsfonts} 

\usepackage{nicefrac}          
\usepackage{microtype}         

\usepackage{booktabs}          

\usepackage{graphicx}          
\usepackage{xcolor}            
\usepackage{subcaption}        
\usepackage{float}             
\usepackage{placeins}          
\usepackage{pdfpages}          

\usepackage[toc,page]{appendix} 

\usepackage{csquotes}          
\usepackage[colorlinks=true]{hyperref}
\usepackage[backend=biber,style=ieee,doi=true,url=false,  style=numeric-comp,
sorting=none]{biblatex}
\AtEveryBibitem{\clearfield{issn}}

\usepackage{hyperref}
\definecolor{myred}{RGB}{160,0,0}
\definecolor{mygreen}{RGB}{0,160,0}
\definecolor{myblue}{RGB}{0,0,160}
\hypersetup{
                colorlinks = true,
                linkcolor = {myred},
                citecolor = {mygreen},
                urlcolor  = {myblue},
}

\graphicspath{ {./images/} }
\newcommand{\sect}[1]{Section~\ref{#1}}
\newcommand{\fig}[1]{Fig.~\ref{#1}}

\newcommand{\equ}[1]{(\ref{#1})}
\newcommand{\appx}[1]{Appendix~\ref{#1}}
\newcommand{\K}{\mathcal{K}}
\newcommand{\U}{\operatorname{V}}
\newcommand{\T}{\operatorname{T}}
\usepackage[normalem]{ulem} 
\numberwithin{equation}{section}

\newtheorem{remark}{Remark}[section]

\newcommand{\RED}[1]{{\textcolor{red}{#1}}}

\newcommand{\rev}[1]{\textcolor{red}{#1}}

\renewcommand{\rev}[1]{#1} 

\usepackage{authblk}

\title{Wave scattering by a transversal defect in a discrete waveguide}

\author[1]{Elena Medvedeva\thanks{\rev{Present address: Mathematics Institute, University of Warwick, Coventry, CV4 7AL, UK. Email address for correspondence:
elena.medvedeva@warwick.ac.uk.}}}
\author[1]{Raphael Assier}
\author[1]{Anastasia Kisil}

\affil[1]{Department of Mathematics of The University of Manchester,

Oxford Road, Manchester, M13 9PL, United Kingdom}

\date{}

\begin{document}
\maketitle


\begin{abstract}
We study wave scattering by a finite transversal strip in a discrete square-lattice waveguide with Dirichlet boundary conditions imposed on the strip and the waveguide walls. The setting is motivated as a discrete analogue of the classical continuous waveguide problem with a screen. The corresponding Wiener–Hopf formulation leads to an equation with a $4 \times 4$ matrix kernel, which reduces to a $2 \times 2$ matrix kernel under some symmetry assumptions.
The prospects for factorising this kernel are discussed, but this approach is not pursued.
Instead, an exact analytical solution is obtained using the pole removal technique. This contrasts with the continuous case, where only approximate solutions are currently available. The reflection and transmission coefficients resulting from an incident duct mode are computed with an accuracy up to $10^{-13}$, showing consistency with theoretical predictions from continuous waveguide theory. In particular, full reflection and zero transmission are recovered as the frequency approaches the cut-off value for the incident mode. Finally, the solution is validated against a numerical computation of the diffraction problem via the Boundary Algebraic Equations method with a tailored lattice Green's function.
\end{abstract}

\input{0_introduction}

\input{2_waveguide_WH}

\input{2_WH_formulation}
\input{4_pole_removal}

\input{5_BAE}

\input{5_0_RT}
\input{5_conclusions}

\section*{Acknowledgementsp}
A.V.K. is supported by a Royal Society Dorothy Hodgkin Research Fellowship and a Dame Kathleen Ollerenshaw Fellowship. E.M. is supported by a President’s Doctoral Scholarship Award of The University of Manchester.
The authors would like to thank the Isaac Newton Institute for Mathematical Sciences, Cambridge, for support and hospitality during the programme `WHT Follow on: the applications, generalisation and implementation of the Wiener-Hopf Method', where work on this paper was undertaken. This programme was supported by EPSRC grant EP/Z000580/1.

\printbibliography

\appendix

\input{append}

\end{document}

%% file: 0_introduction.tex
\section{Introduction}
\label{sect:introduction}

Discrete lattice models have a long history in the analysis of wave propagation and scattering.
One of the most widely used frameworks for modelling discrete systems is the mass-spring model introduced by Slepyan for studying dynamic fracture in discrete structures \cite{Slepyan2002ModelsMechanics}. This simple yet powerful approach has been applied to lattices of masses connected by springs to investigate not only dynamical fracture phenomena in lattice systems \cite{Mishuris2008DynamicsLattice, Nieves2013PropagationLattice, Mishuris2008DynamicalLattice, Colquitt2012TrappingLattice}, but also wave scattering problems, pioneered by Sharma \cite{Sharma2015DiffractionCrack, Sharma2015DiffractionConstraint} and further developed in subsequent works, for example \cite{Nieves2024AnalyticalLoad, Nieves2024InteractionLattice, Medvedeva2024DiffractionMethod}. Similar methods have also been used to study wave propagation in waveguides \cite{Slepyan2022AnSub-structures, Slepyan2022ForcedAttached} and in elastic bodies \cite{Mishuris2020WavesMicrostructure} with embedded microstructures.

Building on Slepyan's work, the theory of wave propagation in discrete waveguides has been studied extensively by Sharma, beginning with \cite{Sharma2017OnPolynomials,Sharma2016WaveStrips}, where dispersion relations and normal modes were derived for infinitely long square- and triangular-lattice waveguides under various boundary conditions on the walls. Some of these results were expressed implicitly through linear combinations of Chebyshev polynomials and were subsequently employed in \cite{Sharma2018OnLattice, Sharma2018OnStructure, Maurya2020WaveCracks}.

The classical problem of scattering by a transversal screen in a continuous waveguide has been studied extensively using the Wiener--Hopf method \cite{Heins1954OnGrating, Weinstein1969Method} among other approaches \cite{Aitken2024OnProblems}; see \cite{Erbas2007ScatteringPlates} for a comprehensive review. This waveguide formulation also arises from the problem of diffraction by an infinite periodic grating by reducing it to a waveguide with quasi-periodic conditions on the walls.
In this work, we consider the discrete analogue of this problem: wave scattering by a finite strip within a discrete waveguide, as shown in \fig{fig:scheme}. 

Similar to the continuous case \cite{Erbas2007ScatteringPlates, Aitken2024OnProblems}, a finite strip in a discrete waveguide leads to a matrix Wiener--Hopf equation. In both discrete and continuous settings, when no specific symmetry is assumed, the kernel is a $4\times4$ matrix, for which no exact factorisation is known. Indeed, factorising matrix kernels is substantially more difficult than factorising scalar kernels \cite{NobleB1958MethodsPDEs, Kisil2021TheMethods, Rogosin2016ConstructiveMatrix-functions}.  However, there are important differences between the discrete and continuous settings, offering a useful framework for investigating how discretisation influences waveguide theory.

In particular, in the continuous formulation, the matrix Wiener--Hopf kernel can be factorised in the Khrapkov--Daniele form \cite{Erbas2007ScatteringPlates, Erbas2002ScatteringStructures} under specific symmetry assumptions. However, as discussed in \sect{sect_matrix_fact}, the matrix kernel arising in the discrete formulation, although seemingly simpler than its continuous counterpart and also belonging to the class of Khrapkov--Daniele matrices \cite{Khrapkov1971CertainForces} under the same symmetry assumption, is nevertheless more difficult to factorise directly.

Moreover, in the general continuous case (no specific symmetry), only approximate solutions can be obtained via the pole removal method \cite{Erbas2007ScatteringPlates, Aitken2024OnProblems}. In contrast, as shown in the present paper, for the discrete problem, the pole removal method leads to a finite system of linear equations and therefore yields an exact solution. 

Additionally, the discrete formulation admits an exact solution through the system of boundary algebraic equations \cite{Poblet-Puig2015SuppressionEquations}, a method that has been applied in related contexts in \cite{Sharma2015Near-tipCrack, Sharma2015Near-tipConstraint, Medvedeva2024DiffractionMethod}. An analogy between the continuous and discrete Wiener--Hopf formulations of diffraction problems was also discussed in \cite{Korolkov2025OnProblems}, where waveguide problems were considered in particular.

In summary, analysing the discrete waveguide formulation enables an assessment of the applicability of mathematical techniques developed for the continuous case, as in \cite{Medvedeva2024DiffractionMethod,Livasov2019NumericalZone}, and provides a way to examine whether the theoretical predictions for reflection and transmission coefficients obtained in the continuous setting \cite{Shanin2017DiffractionWaveguideb} are preserved in the discrete analogue.

This paper is organised as follows. In \sect{sect:waveguide} we formulate the scattering problem for a discrete waveguide with a transversal screen and discuss the structure of the wavefield inside the waveguide. In \sect{sect_WH_formulation} we formulate the matrix Wiener--Hopf problem for the general and symmetric cases, followed by a discussion on the prospects of matrix kernel factorisation in \sect{sect_matrix_fact}. In \sect{sect_pole_removal} we derive a solution to the Wiener--Hopf problem using the pole removal technique, and in \sect{eq_bae} we derive an expression for a tailored lattice Green's function for a discrete waveguide and formulate a solution in terms of a system of boundary algebraic equations. In \sect{sect:future}, we analyse the obtained reflection and transmission coefficients. Finally, conclusions are drawn in \sect{sect_conclusion}.

%% file: 2_waveguide_WH.tex
\section{Discrete waveguide with a transversal screen}
\label{sect:waveguide}
\subsection{Geometry of the problem}
\begin{figure}[H]
    \centering
    \includegraphics[width=.45
    \textwidth]{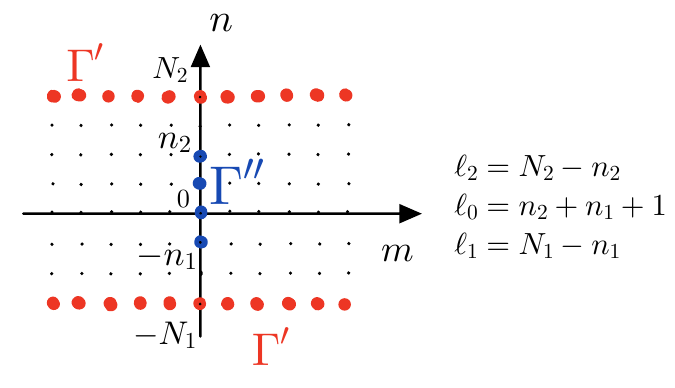} 
    \caption{A discrete waveguide of a square lattice with a transversal Dirichlet screen and Dirichlet 
    waveguide walls, $u^\text{tot}_{m,n} = 0$ for $(m,n)\in\Gamma',\Gamma''$}.
    \label{fig:scheme}
\end{figure}

We consider an incident duct mode plane wave within an infinite discrete square-lattice waveguide containing a transversal defect (\fig{fig:scheme}). The lattice supports the propagation of anti-plane waves. 

The total displacement field $u^\text{tot}_{m,n}$ in the waveguide satisfies the discrete Helmholtz equation:

\begin{equation}
    \Delta u^\text{tot}_{m,n} + \Omega^2 u^\text{tot}_{m,n} = 0, \quad \{m,n\}\in(\mathbb{Z}\times\mathbb{Z}_N) \backslash (\Gamma''\cup\Gamma'),
    \label{eq_Helmholtz_discrete}
\end{equation}
where
\begin{equation}
    \Delta u^{\text{tot}}_{m,n} = u^{\text{tot}}_{m+1,n} + u^{\text{tot}}_{m-1,n} + u^{\text{tot}}_{m,n+1} + u^{\text{tot}}_{m,n-1} - 4 u^{\text{tot}}_{m,n}.
    \label{eq_laplacian_5point}
\end{equation}
is the five-point discrete Laplacian, $\Omega$ is the lattice frequency, $\mathbb{Z}_N \equiv \{\,n\in\mathbb{Z} : -N_1 \leq n \leq N_2\,\}$, and $\Gamma'$ and $\Gamma''$ are the sets of nodes defining the waveguide walls and the transversal screen:
\begin{align}
    &\Gamma'=\{m\in\mathbb{Z}; \ n=-N_1 \cup n=N_2\},
    \\
   & \Gamma'' = \{m=0; -n_1\leq n\leq n_2\}.
   \label{eq_gamma_dashdash}
\end{align}

We set Dirichlet boundary conditions for the total field on the screen and the waveguide walls:
\begin{align}
    &u^\text{tot}_{m,n} = 0, \qquad \{m,n\}\in\Gamma'';
    \label{eq_strip_bc}
    \\
    &u^\text{tot}_{m,n} = 0, \qquad \{m,n\}\in\Gamma'.\label{eq_bc_walls}
\end{align}

The total field is defined as a sum of the known incident field $u^\text{in}_{m,n} $ and the unknown scattered field $u^\text{sc}_{m,n}$:
\begin{equation}
     u^\text{tot}_{m,n} =  u^\text{in}_{m,n} +  u^\text{sc}_{m,n}. 
     \label{eq_tot_sc_in}
\end{equation}

\subsection{Dispersion relation}

We will seek a solution to \equ{eq_Helmholtz_discrete}, \equ{eq_strip_bc}-\equ{eq_tot_sc_in} as a combination of  waves of the form:
\begin{equation}
   u_{m,n} = x^m\ \left(y^{n}-y^{-n}\right) = x^m\operatorname{s}(n), \label{eq_sol_form_tot}
\end{equation} 
where we defined the `sine' and, for future convenience, the
`cosine' functions as:
\begin{align}
    \operatorname{s}(n) := y^{n} - y^{-n}, \qquad
    \operatorname{c}(n) := y^{n} + y^{-n}. 
    \label{eq_s_def}
\end{align}

Note that, for brevity, we do not explicitly indicate in the notation that $\operatorname{s}(n)$ and $\operatorname{c}(n)$ depend on $y$ (and hence on $x$).

We assume that $|x|\leqslant1,\ |y|\leqslant1$ in \equ{eq_sol_form_tot}, and that $x,\ y$
satisfy the lattice dispersion relation:
\begin{equation}
    \Omega^2 - 4 + x + x^{-1} + y + y^{-1}= 0.
    \label{eq_disp_rel_waveguide}
\end{equation}

Note that only the waves satisfying the boundary conditions on the waveguide walls \equ{eq_bc_walls} may exist in the waveguide \cite{Sharma2017OnPolynomials,Sharma2016WaveStrips}.  Each such wave can be written as a superposition of waveguide modes $u_{m,n}^j$ defined by
\begin{equation}
    u^j_{m,n} = 
     x_j^m \operatorname{s}_j(n+N_1), \qquad \operatorname{s}_{j}(n) := \left[y_j^{n} - y_j^{-n}\right] 
     ,
    \label{eq_mode_form}
\end{equation}
where $x_j$ and $y_j$ are given by
\begin{equation}
    x_j=e^{iK_j}, \quad y_j = e^{i\theta_j},\quad \theta_j = \frac{j\pi}{N},
    \label{eq_theta}
\end{equation}
for some real Bloch wavenumber $K_j$, whose values are restricted to the Brillouin zone $[-\pi,\pi]$, and
\begin{equation}
    N=N_1+N_2. 
    \label{eq_N_def}
\end{equation}
Therefore, each mode has a sine shape with a complex amplitude:
\begin{equation*}
    u^j_{m,n} = 2 i\ x^m_j \sin{\left[\theta_j (n+N_1)\right]},
\end{equation*}
containing an integer number of half-periods between the walls. The piston mode (plane waves) of the form $x_j^m$ cannot propagate in the Dirichlet waveguide due to the boundary conditions on the walls. Because of periodicity, there are $N-1$ physically meaningful values of $j$: $j \in \{1, \dots, N-1\}$. This contrasts with continuous waveguides, which may support infinitely many modes.

Substituting \equ{eq_mode_form} into the discrete Helmholtz equation \equ{eq_Helmholtz_discrete} yields the lattice dispersion relation \equ{eq_disp_rel_waveguide} for the $j$-th waveguide mode:
\begin{equation}
    \Omega^2 - 4 + x_j + x_j^{-1} + y_j + y_j^{-1}=\Omega^2 - 4 + 2\cos K_j + 2\cos\theta_j= 0.
    \label{eq_disp_rel_waveguide_in}
\end{equation}
From \equ{eq_theta}, $\theta_j \in [0,\pi]$, and the dispersion relation \equ{eq_disp_rel_waveguide_in} implies that $\Omega$ lies within $\left[ 0, 2\sqrt{2} \right]$ for the lattice to support plane-wave propagation. 
\rev{\begin{remark}For the infinite square lattice, the values $\Omega = 0$, $\Omega = 2$ and $\Omega = 2 \sqrt{2}$ are resonant degeneracies, as explained in \cite{Martin2006DiscreteLattice, Vanel2016AsymptoticsFunctions, Shanin2024DoubleStructures}. However, in the waveguide considered here, which has a finite cross section and Dirichlet boundary conditions on its walls, we observe no resonance at $\Omega=2,$ and no mode can propagate at the interval endpoints $\Omega=\{0,2\sqrt{2}\},$ as discussed in the next paragraph. Thus, these degeneracies do not affect our further analysis.
\end{remark}}
\begin{figure}[H]
    \centering
    \includegraphics[width=.45
    \textwidth]{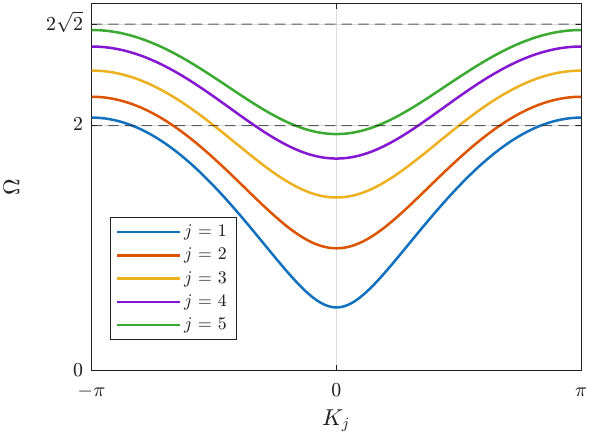} 
    \caption{\rev{Dispersion curves for modes $j\in\{1,\ldots,N-1\}$ for the waveguide with the size of a cross-section $N=6$ with the Dirichlet boundary condition imposed on the walls.}}
    \label{fig_dispersion}
\end{figure}
\rev{
An example of the dispersion curves for the Dirichlet waveguide with $N=6$ is shown in \fig{fig_dispersion}.
Note that the $j$-th mode $u^j_{m,n}$ is propagating when the lattice frequency $\Omega$ lies in a mode propagation band:
\begin{equation}
    \Omega_j^{\text{min}}\leqslant\Omega\leqslant\Omega_j^{\text{max}},
    \label{eq_omega_interval}
\end{equation}
where $\Omega_j^{\text{min}}$ and $\Omega_j^{\text{max}}$ for $j\in\{1,\ldots,N-1\}$ are obtained from $K_j=0$ and $K_j=\pi$ in \equ{eq_disp_rel_waveguide_in} respectively:
\begin{align}
    &\Omega_j^{\text{min}}=\sqrt{2(1-\cos{\theta_j})}, \qquad \Omega_j^{\text{min}}\in(0,2), \label{eq_omega_crit_def}
    \\
    &\Omega_j^{\text{max}}=\sqrt{2(3-\cos{\theta_j})}, \qquad \Omega_j^{\text{max}}\in(2,2\sqrt{2}). \label{eq_omega_crit_max_def}
\end{align}
 For $\Omega$ outside the interval \equ{eq_omega_interval}, the $j$-th mode is evanescent, does not carry any energy, decays rapidly and hence does not propagate in the waveguide. In particular, waves with frequency below the first minimal critical frequency $\Omega_1^{\text{min}}$ or above $\Omega_{N-1}^{\text{max}}$ cannot propagate in a duct with Dirichlet boundary conditions on its walls, resulting in upper and lower band gaps.\footnote{\rev{In contrast, in a continuous waveguide each mode only has a lower cut-off frequency, above which it remains propagating \cite{Shanin2017DiffractionWaveguideb}. The upper cut-off frequency in the discrete case arises from the bounded dispersion relation associated with the lattice periodicity.}}}
\rev{\begin{remark}
Note that, if we take the cross section of the waveguide to be infinitely large, \equ{eq_omega_crit_def}-\equ{eq_omega_crit_max_def} have the following limits as $N\to\infty$:
\begin{align*}
    &\Omega_1^{\text{min}}\to0, && \Omega_1^{\text{max}}\to2,
    \\
    &\Omega_{N-1}^{\text{min}}\to2, && \Omega_{N-1}^{\text{max}}\to2\sqrt{2}.
\end{align*}
Hence, the dispersion relation for the waveguide transforms into the dispersion relation of the infinite square lattice, where the dispersion curve for each mode is a cross-section of the dispersion surface of the square lattice (see, for example, \cite{Sharma2015Near-tipConstraint, Vanel2016AsymptoticsFunctions}). Since the waveguide with an infinite cross section supports an infinite number of propagating modes, $\theta_j$ transforms into the Bloch wavenumber $\Theta$, which takes values in the continuous spectrum $[0,\pi]$. Physically, this represents all plane waves propagating in the positive direction of the $m$-axis. 
\end{remark}}


Thus, we define the incident field $u^\text{in}_{m,n}$ in \equ{eq_tot_sc_in} as a propagating waveguide mode of the form \equ{eq_mode_form}:
\begin{equation}
    u^\text{in}_{m,n} = u^p_{m,n}=
     x^m_p \operatorname{s}_p(n+N_1), 
    \label{eq_u_in_duct}
\end{equation}
for some $p\in\{1,\ldots,N-1\}$ such that  \rev{its lattice frequency $\Omega$ satisfies $\Omega_p^{\text{min}}<\Omega<\Omega_p^{\text{max}}$}. The function $\operatorname{s}_p$ and the associated $y_p$ are defined as in  \equ{eq_mode_form}-\equ{eq_theta}.

\subsection{Boundary-value problem for the waveguide}
\label{sect_waveguide_prob_form}

For scattering problems in infinite domains, for the problem
to be well posed, one must specify the \textit{radiation condition} at infinity. We will do this through the \textit{limiting
absorption principle}: if $\Omega$ is assumed to have a small positive
imaginary part $\varepsilon_{\Omega}$, we set the condition by imposing that the
resulting scattered field $u^\text{sc}_{m,n}$ must be exponentially decaying at
infinity. The true physical solution is then understood as the limit of these
admissible solutions when $\varepsilon_{\Omega}\rightarrow 0^+$.

\rev{
From the boundary condition on the screen, together with uniqueness of the radiating solution specified through the limiting absorption principle, we may show (see \appx{sect:app_symmetry}) that the scattered field is symmetric in $m$:
}
\begin{equation}
    u^\text{sc}_{m,n}=u^\text{sc}_{-m,n}.
    \label{eq_field_symmetry}
\end{equation}

Then from \equ{eq_Helmholtz_discrete} and \equ{eq_field_symmetry} it follows for $m=0,\ -N_1<n<\RED{-}n_1$ and $m=0,\ n_2<n<N_\RED{2}$:
\begin{equation}
    \frac{1}{2}(u^\text{sc}_{0,n+1}+u^\text{sc}_{0,n-1})+u^\text{sc}_{1,n}+\left(\frac{\Omega^2}{2}-2\right)u^\text{sc}_{0,n} =0.
    \label{eq_BC_newmann}
\end{equation}

Note also that the incident field \equ{eq_u_in_duct} is defined such that:
\begin{equation}
   u^\text{in}_{m,-N_1}= u^\text{in}_{m,N_2}\equiv0.
   \label{eq_walls_boundary_in}
\end{equation}

Since the problem is symmetric with respect to $m=0$, we only need to consider the right half of the waveguide where $m\geqslant0$. Then, from the Helmholtz equation \equ{eq_Helmholtz_discrete}, the boundary conditions on the walls \equ{eq_bc_walls} and \equ{eq_walls_boundary_in}, and the boundary conditions \equ{eq_strip_bc} and \equ{eq_BC_newmann}, we formulate the mixed boundary-value problem for the scattered field:
\begin{align}
&\Delta u^\text{sc}_{m,n} + \Omega^2 u^\text{sc}_{m,n} = 0, && m>0,\ -N_1<n<N_2;
\label{eq_Helmholtz_discrete_sc}
\\
&u^\text{sc}_{m,-N_1} = u^\text{sc}_{m,N_2} = 0, && m\geqslant0;
\label{eq_walls_boundary}
\\
&u^\text{sc}_{0,n} = - u^\text{in}_{0,n}, && -n_1\leqslant n\leqslant n_2;
\label{eq_BC_symmetric1}
\\
&2u^\text{sc}_{1,n} + u^\text{sc}_{0,n+1} + u^\text{sc}_{0,n-1} +(\Omega^2-4) u^\text{sc}_{0,n} = 0, && -N_1<n<-n_1 \text{ or } n_2<n<N_2.
\label{eq_BC_symmetric2}
\end{align}

%% file: 2_WH_formulation.tex
\FloatBarrier
\section{Wiener--Hopf formulation}
\label{sect_WH_formulation}
\subsection{Odd and even discrete Fourier transforms}
In order to find the analytical solution, we will derive a Wiener--Hopf formulation for the discrete waveguide with a transversal screen in a similar manner to \cite{Erbas2007ScatteringPlates}, where the Wiener--Hopf equation is written for the continuous waveguide with periodic boundary conditions on the walls.

In the discrete waveguide problem, to pass to the spectral domain and derive a Wiener--Hopf equation, the geometry allows only a Fourier transform along the $m$ axis (see \fig{fig:scheme}), which, unlike the usual formulations \cite{Sharma2015DiffractionConstraint, Sharma2015Near-tipCrack, Medvedeva2024DiffractionMethod}, is perpendicular to the strip. For this reason, in the continuous analogue, the authors of \cite{Erbas2007ScatteringPlates} used sine or cosine transforms, motivated by the boundary conditions on the screen and the adjacent gap.

In the discrete case, we therefore introduce the discrete analogues of the sine and cosine transforms. We first define the discrete Fourier transform of $u^\text{sc}_{m,n}$ as
\begin{equation}
    U(x,n) = \sum_{m=-\infty}^{\infty} u^\text{sc}_{m,n}x^{-m} =U_-(x,n) + U_+(x,n),
  \label{eq_discrete_Fourier_transform}
\end{equation}
where 
\begin{align*}
    U_+(x,n) = \sum_{m=0}^{\infty} u^\text{sc}_{m,n} x^{-m} \quad \text{and} \quad U_-(x,n) = \sum_{m=-\infty}^{-1} u^\text{sc}_{m,n} x^{-m}
\end{align*}
are the half-transforms.
As discussed above, $\Omega$ has a small imaginary part $\varepsilon_\Omega$ due to the limiting absorption principle, so it follows from the dispersion relation and symmetry that 
\[
|u^\text{sc}_{m,n}| < e^{-\varepsilon |m|},
\]
where $\varepsilon$ is the minimal imaginary part of all $K_j$, arising from $\varepsilon_\Omega$ in the dispersion relation.  
Thus, the functions $U_-(x,n)$ and $U_+(x,n)$ are analytic for $|x| < e^{\varepsilon}$ and $|x| > e^{-\varepsilon}$, respectively. We will further refer to these as the regions inside and outside the unit circle $|x|=1$. Hence, \equ{eq_discrete_Fourier_transform} is analytic in the annulus 
\[
\mathcal{A} : \{ e^{-\varepsilon} < |x| < e^{\varepsilon} \},
\]
which includes the unit circle.

The inverse transform to recover the physical field $u^\text{sc}_{m,n}$ is
\begin{equation}
    u^\text{sc}_{m,n}=\frac{1}{2 \pi i} \oint_{\mathcal{C}} U(x,n) x^{m-1} d x, \quad m \in \mathbb{Z},
    \label{eq_discr_inv_fourier_HP}
\end{equation}
where $\mathcal{C}$ is a closed counter-clockwise contour, which lies in the annulus of analyticity $\mathcal{A}$ of $U(x,n)$ and encircles the origin.
Note that, due to the symmetry \equ{eq_field_symmetry},
\begin{equation}
    U_-(x,n) = U_+(x^{-1},n)-u^\text{sc}_{0,n}. 
    \label{eq_U_symmetry}
\end{equation}
Therefore, it is sufficient to find $U_+(x,n)$ and consequently $u^\text{sc}_{0,n}$ to determine the full transform $U(x,n)$.

We will then define the discrete analogues of the sine and cosine transforms -- the \textit{odd} $U_{(o)}(x,n)$ and \textit{even} $U_{(e)}(x,n)$ transforms:
\begin{alignat}{2}
    &U_{(e)}(x,n) &&= \sum_{m=0}^{\infty} u^\text{sc}_{m,n} (x^{m} + x^{-m})= \Phi(x,n) + u^\text{sc}_{0,n}, \label{eq_transform_even} \\
    &U_{(o)}(x,n) &&= \sum_{m=0}^{\infty} u^\text{sc}_{m,n} (x^{m} - x^{-m}) = \Psi(x,n) + u^\text{sc}_{0,n}, \label{eq_transform_odd}
\end{alignat}
where 
the functions $\Phi(x,n)$ and $\Psi(x,n)$ are defined as:
\begin{align}
    \Phi(x,n) & = U_-(x,n) + U_+(x,n), \label{eq_phi_definition}\\
    \Psi(x,n) &= U_-(x,n) - U_+(x,n).
    \label{eq_psi_definition}
\end{align}
Note also that by \equ{eq_discrete_Fourier_transform} and \equ{eq_phi_definition} it follows that $\Phi(x,n)\equiv U(x,n)$.

The transforms defined by \equ{eq_transform_even}-\equ{eq_transform_odd} are analytic in the annulus of analyticity $\mathcal{A}.$ Integrating \equ{eq_transform_even} and \equ{eq_transform_odd} over $\mathcal{C}$, exchanging the sum and integral symbols,  and using the fact that 
$$\frac{1}{2\pi i}\oint_\mathcal{C} x^{-m-1} dx = \delta_{m,0},$$ 
where $\delta$ is the usual Kronecker delta, we obtain the inverse transforms of \equ{eq_transform_even} and \equ{eq_transform_odd} as:
\begin{alignat}{2}
    &u^\text{sc}_{m,n}=\frac{1}{2 \pi i} \oint_{\mathcal{C}} U_{(e)}(x,n) x^{-m-1} d x - u^\text{sc}_{0,n}\delta_{m,0}, \quad m \in \mathbb{Z}^+_0; \label{eq_transform_even_inv0}\\
    &u^\text{sc}_{m,n}=\frac{1}{2 \pi i} \oint_{\mathcal{C}} U_{(o)}(x,n)x^{-m-1} d x + u^\text{sc}_{0,n}\delta_{m,0}, \quad m \in \mathbb{Z}^+_0, \label{eq_transform_odd_inv0}
\end{alignat}
where $\mathbb{Z}^+_0=\{m\in\mathbb{Z}:m\geqslant0\}$, $\mathcal{C}$ is a closed counter-clockwise contour, which lies within the annulus $\mathcal{A}$, and  $\delta_{i,j}$ is the Kronecker delta, defined as
\begin{equation}
    \delta_{i,j}=
    \begin{cases}
        1,\quad i=j, \\
        0,\quad i\neq j.
    \end{cases}
    \label{eq_kronecker_def}
\end{equation}

Using the definitions \equ{eq_transform_even}-\equ{eq_transform_odd}, we may rewrite \equ{eq_transform_even_inv0}-\equ{eq_transform_odd_inv0} as:
\begin{alignat}{2}
    &u^\text{sc}_{m,n}=\frac{1}{2 \pi i} \oint_{\mathcal{C}} \Phi(x,n) x^{m-1} d x, \quad m \in \mathbb{Z}^+_0; \label{eq_transform_even_inv}\\
    &u^\text{sc}_{m,n}=\frac{1}{2 \pi i} \oint_{\mathcal{C}} \Psi(x,n)x^{m-1} d x + 2u^\text{sc}_{0,n}\delta_{m,0}, \quad m \in \mathbb{Z}^+_0. \label{eq_transform_odd_inv}
\end{alignat}
Note that we are allowed to change the sign in front of $m$ in $x^{-m-1}$ in \equ{eq_transform_even_inv}-\equ{eq_transform_odd_inv} due to the symmetry of the problem \equ{eq_field_symmetry}.

\subsection{Wiener--Hopf problem formulation}


Similarly to the continuous analogue \cite{Erbas2007ScatteringPlates}, motivated by the boundary conditions \equ{eq_BC_symmetric1}-\equ{eq_BC_symmetric2} we apply 
the even transform \equ{eq_transform_even} to the Helmholtz equation \equ{eq_Helmholtz_discrete_sc} for 
$-N_1<n<-n_1$ or $n_2<n<N_2$
, we get:
\begin{equation}
    \Lambda(x) \Phi(x,n) + \Phi(x,n+1) + \Phi(x,n-1)
    +2u^\text{sc}_{1,n}+[\Omega^2-4] u^\text{sc}_{0,n} + u^\text{sc}_{0,n+1} + u^\text{sc}_{0,n-1} =0,
    \label{eq_helmholts_odd}
\end{equation}
where $\Phi(x,n)$ is defined in \equ{eq_phi_definition} and 
\begin{equation}
\Lambda(x) = \Omega^2-4 +x+x^{-1}.
    \label{eq_Lambda_def}
\end{equation}
Applying  the odd transform \equ{eq_transform_odd} to the Helmholtz equation \equ{eq_Helmholtz_discrete_sc} for 
$-n_1\leqslant n\leqslant n_2$, we get:
\begin{equation}
    \Lambda(x) \Psi(x,n) + \Psi(x,n+1) + \Psi(x,n-1)+[\Omega^2-4 + 2x] u^\text{sc}_{0,n} + u^\text{sc}_{0,n+1} + u^\text{sc}_{0,n-1} = 0, \label{eq_helmholts_even}
\end{equation}
where $\Psi(x,n)$ is defined in \equ{eq_psi_definition}.
Note that to obtain \equ{eq_helmholts_odd}-\equ{eq_helmholts_even} we used the following relations:
\begin{align*}
    &\sum_{m=0}^{\infty} u^\text{sc}_{m+1,n} x^{m} = x^{-1} U_-(x,n),& 
    &\sum_{m=0}^{\infty} u^\text{sc}_{m-1,n} x^{m} = x U_-(x,n) + x u^\text{sc}_{0,n} + u^\text{sc}_{1,n}, 
    \\
    &\sum_{m=0}^{\infty} u^\text{sc}_{m-1,n} x^{-m} = x^{-1} U_+(x,n) + u^\text{sc}_{1,n},& 
    &\sum_{m=0}^{\infty} u^\text{sc}_{m+1,n} x^{-m} = x U_+(x,n) - x u^\text{sc}_{0,n}.
\end{align*}

Then from \equ{eq_helmholts_odd}-\equ{eq_helmholts_even} and the boundary conditions \equ{eq_BC_symmetric1}-\equ{eq_BC_symmetric2} it follows that
\begin{align}
&\Lambda(x) \Phi(x,n) + \Phi(x,n+1) + \Phi(x,n-1)=0, \quad &&-N_1<n<-n_1\quad \text{or}\quad n_2<n<N_2;
\label{eq_diff_eqs1}
\\
    &\Lambda(x) \Psi(x,n) + \Psi(x,n+1) + \Psi(x,n-1) = r(x,n), \quad &&-n_1\leqslant n\leqslant n_2,
\label{eq_diff_eqs2}
\end{align}
where from \equ{eq_BC_symmetric1} and \equ{eq_helmholts_odd}-\equ{eq_helmholts_even} we define the function $r(x,n),$ known for  $-n_1<n< n_2,$ as:
\begin{equation}
    r(x,n) = \left[\Lambda(x)+y_p + y_p^{-1} + \left(x-x^{-1}\right) \right]\operatorname{s}_p(n+N_1). \label{eq_r_def_gen}
\end{equation}
For $n=-n_1$ and $n=n_2$, $r(x,n)$ in \equ{eq_diff_eqs2} is known up to the two unknown constants $u^\text{sc}_{0,-n_1-1}$ and $u^\text{sc}_{0,n_2+1}$:
\begin{alignat}{3}
    & r(x,-n_1) & &= \left[\Lambda(x) + \left(x-x^{-1}\right) \right]\operatorname{s}_p(n_1^*) + \operatorname{s}_p(n_1^*+1) - u^\text{sc}_{0,-n_1-1}, \label{eq_r_def01}
    \\
    & r(x,n_2) & &= \left[\Lambda(x) + \left(x-x^{-1}\right) \right]\operatorname{s}_p(n_2^*) +\operatorname{s}_p(n_2^*-1) - u^\text{sc}_{0,n_2+1},
\label{eq_r_def02}
\end{alignat}
where $n^*_1 = -n_1+N_1,\ n^*_2 = n_2+N_1$.

Equations \equ{eq_diff_eqs1}-\equ{eq_diff_eqs2} are a homogeneous and an inhomogeneous difference equation, respectively. From the theory of difference equations \cite{Levy1961FiniteEquations}, we seek solutions to \equ{eq_diff_eqs1}-\equ{eq_diff_eqs2} using the ansatz
\begin{alignat}{2}
     &\Phi(x,n) = A_1(x)\ y^n + B_1(x)\ y^{-n}, \quad &&-N_1<n<-n_1;
     \label{eq_ansatz_1}
     \\
    &\Phi(x,n) = A_2(x)\ y^n + B_2(x)\ y^{-n}, \quad &&n_2<n<N_2;
    \label{eq_ansatz_2}
    \\
    &\Psi(x,n) = C(x)\ y^n + D(x)\ y^{-n} + R(x,n), \quad &&-n_1\leqslant n\leqslant n_2,
    \label{eq_ansatz_b}
\end{alignat}
where $A_1(x),\ B_1(x),\ A_2(x),\ B_2(x),\ C(x),\ D(x)$ are unknown functions, and the function $y(x)$ in \equ{eq_ansatz_1}-\equ{eq_ansatz_b} obeys the dispersion relation \equ{eq_disp_rel_waveguide} and
is defined as:
\begin{equation}
    y(x) = -\frac{\Lambda(x)}{2}+\frac{\sqrt{\Lambda(x)^2-4}}{2},
    \qquad  
    (y(x))^{-1}=-\frac{\Lambda(x)}{2}-\frac{\sqrt{\Lambda(x)^2-4}}{2},
    \label{eq_y_definition}
\end{equation}
where we choose the branch of the square root such that $\operatorname{Im}(\sqrt{\Lambda^2-4})>0$, and therefore $|y(x)|\leqslant 1$.

The function $R(x,n)$ in \equ{eq_ansatz_b} is a particular solution of the inhomogeneous equation \equ{eq_diff_eqs2}, 
which is defined from \equ{eq_r_def_gen} as:
\begin{equation}
R(x,n) = \frac{r(x,n)}{\Lambda(x)+y_p+y_p^{-1}}=\Upsilon(x)\operatorname{s}_p(n+N_1),
\label{eq_R}
\end{equation}
where
\begin{equation}
    \Upsilon(x) 
= 1+\Pi(x),
\qquad \Pi(x)=\frac{x^2-1}{(x-x_p)(x-x_p^{-1})},
\label{eq_upsilon_pi}
\end{equation}
and the function $\Pi(x)$ is found from the dispersion relation for the incident field \equ{eq_disp_rel_waveguide_in} and the definition of $\Lambda$ \equ{eq_Lambda_def}, where $x_p$  is defined such that $|x_p|\leqslant1.$ 



We can also rewrite \equ{eq_r_def01}-\equ{eq_r_def02} in terms of $\Upsilon(x)$:
\begin{alignat}{3}
    & r(x,-n_1) & &= \Upsilon(x)\left[\Lambda(x)\operatorname{s}_p(n_1^*) + \operatorname{s}_p(n_1^*+1) + \operatorname{s}_p(n_1^*-1)\right] - (\operatorname{s}_p(n_1^*-1) + u^\text{sc}_{0,-n_1-1}), \label{eq_r_def0_1}
    \\
    & r(x,n_2) & &= \Upsilon(x)\left[\Lambda(x) \operatorname{s}_p(n_2^*) + \operatorname{s}_p(n_2^*-1) + \operatorname{s}_p(n_2^*+1)\right] - (\operatorname{s}_p(n_2^*+1) + u^\text{sc}_{0,n_2+1}). \label{eq_r_def0_2}
\end{alignat}

The boundary conditions \equ{eq_walls_boundary} on the waveguide walls imply that we have 
\begin{align}
    \Phi(x,-N_1) =0, 
    \qquad \Phi(x,N_2 )=0. \label{eq_bc_spectral}
\end{align}

Hence, writing \equ{eq_ansatz_1}-\equ{eq_ansatz_2} for $n=-N_1+1$ and $n=N_2-1$, respectively, and using  \equ{eq_bc_spectral}, we can eliminate the unknowns $B_1(x)$ and $A_2(x)$ from \equ{eq_ansatz_1}-\equ{eq_ansatz_2}:
\begin{align}
    &\Phi(x,n) = A_1(x)\operatorname{s}(N_1+n), \qquad &&-N_1<n<-n_1; \label{eq_ansatz_1A}
    \\
    &\Phi(x,n) = B_2(x)\operatorname{s}(N_2-n), \qquad &&\quad n_2<n<N_2.  \label{eq_ansatz_2A}
\end{align}
Then, if we express $A_1(x)$ and $B_2(x)$ from \equ{eq_ansatz_1A}-\equ{eq_ansatz_2A} written for $n=-n_1-1,\ n=n_2+1$, respectively, we arrive at:
\begin{align}
    &\Phi(x,n) = \Phi(x,-n_1-1)\frac{\operatorname{s}(N_1+n)}{\operatorname{s}(N_1-n_1-1)},  &&-N_1<n<-n_1; \label{eq_ansatz_1A2}
    \\
    &\Phi(x,n) = \Phi(x,n_2+1)\frac{\operatorname{s}(N_2-n)}{\operatorname{s}(N_2-n_2-1)} ,  &&n_2<n<N_2.  \label{eq_ansatz_2A2}
\end{align}


Writing equation \equ{eq_diff_eqs1} for $n=-n_1-1$ and $n=n_2+1$, which are the rows adjacent to the screen's edge, and using \equ{eq_ansatz_1A2}-\equ{eq_ansatz_2A2}, 
we find:
\begin{alignat}{7}
    &\left[\Lambda(x) + \frac{\operatorname{s}(\ell_1-2)}{\operatorname{s}(\ell_1-1)}\right]\Phi(x,-n_1-1) & &+ & &\Phi(x,-n_1) & &= 0,  \label{eq_eq_sys1}\\
    & \left[\Lambda(x) + \frac{\operatorname{s}(\ell_2-2)}{\operatorname{s}(\ell_2-1)}\right] \Phi(x,n_2+1) & &+ & &\Phi(x,n_2) & &=0, \label{eq_eq_sys2}
\end{alignat}
where $\Phi(x,-n_1)$ and $\Phi(x,n_2)$ are defined by \equ{eq_phi_definition},
and $\ell_1,\ \ell_2$ are the lengths of the gaps below and above the strip, defined as:
\begin{align}
    \ell_1 &:= N_1 - n_1,\qquad \ell_2 := N_2 - n_2, \label{eq_ell1_ell2_def}
\end{align}
where we restrict $\ell_1\geqslant2$ and $\ell_2\geqslant2$.

Note that by definition of $\operatorname{s}(n)$ \equ{eq_s_def} and $\Lambda(x)$ \equ{eq_Lambda_def} we may show that for any integer $\ell\geqslant2$ we have 
\begin{equation}
    \Lambda(x)\operatorname{s}(\ell-1)+\operatorname{s}(\ell-2) = -\operatorname{s}(\ell).  \label{eq_l_property}
\end{equation}
Then, multiplying \equ{eq_eq_sys1}-\equ{eq_eq_sys2} by $\operatorname{s}(\ell_1-1)$ and $\operatorname{s}(\ell_2-1)$, respectively, and using \equ{eq_l_property}, we obtain the following pair of equations:
\begin{alignat}{7}
    &-\operatorname{s}(\ell_1)\Phi(x,-n_1-1) &&+ &&\operatorname{s}(\ell_1-1)\Phi(x,-n_1) &&=0,  \label{eq_pair1_1}\\
    & -\operatorname{s}(\ell_2)\Phi(x,n_2+1) &&+  &&\operatorname{s}(\ell_2-1)\Phi(x,n_2) &&=0. \label{eq_pair1_2}
\end{alignat}

Similarly, we can evaluate \equ{eq_diff_eqs2} for $n=-n_1$ and $n=n_2$ and obtain $C(x), \ D(x)$ from the resulting system as:
\begin{align}
    C(x) = &-\left[\Psi(x,-n_1)-R(x,-n_1)\right] \frac{y^{-n_2}}{\operatorname{s}(n_1+n_2)}+
    \left[\Psi(x,n_2)-R(x,n_2)\right]\frac{y^{n_1}}{\operatorname{s}(n_1+n_2)}, \label{eq_C_expr}
    \\
    D(x) = 
    &+
    \left[\Psi(x,-n_1)-R(x,-n_1)\right] \frac       {y^{n_2}}{\operatorname{s}(n_1+n_2)}-\left[\Psi(x,n_2)-R(x,n_2)\right]\frac{y^{-n_1}}{\operatorname{s}(n_1+n_2)}. \label{eq_D_expr}
\end{align}

Then, substituting \equ{eq_C_expr}-\equ{eq_D_expr} into \equ{eq_ansatz_b}, inputting the resulting expression for $\Psi$ in \equ{eq_diff_eqs2} for $n=-n_1$ and $n=n_2,$ and using \equ{eq_l_property}, we find a second pair of equations:
\begin{align}
    &-\frac{\operatorname{s}(\ell_0)}{\operatorname{s}(1)} \Psi(x,-n_1) + \Psi(x,n_2) +
    \frac{\operatorname{s}(\ell_0-1)}{\operatorname{s}(1)}\Psi(x,-n_1-1) = R_1(x),
    \label{eq_pair2_1}
    \\
    &-\frac{\operatorname{s}(\ell_0)}{\operatorname{s}(1)} \Psi(x,n_2) + \Psi(x,-n_1) 
    + \frac{\operatorname{s}(\ell_0-1)}{\operatorname{s}(1)} \Psi(x,n_2+1)  = R_2(x),
    \label{eq_pair2_2}
\end{align}
where, using \equ{eq_N_def}, \equ{eq_R}-\equ{eq_r_def0_2} and \equ{eq_ell1_ell2_def} after some simplification we defined
\begin{align}
    R_1(x) &= \Upsilon(x)\left[\operatorname{s}_p(N-\ell_2) - \frac{\operatorname{s}(\ell_0)}{\operatorname{s}(1)} \operatorname{s}_p(\ell_1)\right] + 
    \frac{\operatorname{s}(\ell_0-1)}{\operatorname{s}(1)}\left[\Pi(x)\operatorname{s}_p(\ell_1-1)
     - u^\text{sc}_{0,-n_1-1}\right],
     \label{eq_R1_x3} 
      \\
     R_2(x) &= \Upsilon(x)\left[ \operatorname{s}_p(\ell_1) - \frac{\operatorname{s}(\ell_0)}{\operatorname{s}(1)}  \operatorname{s}_p(N-\ell_2) \right] +
    \frac{\operatorname{s}(\ell_0-1)}{\operatorname{s}(1)} 
     \left[\Pi(x)\operatorname{s}_p(N-(\ell_2-1))-u^\text{sc}_{0,n_2+1}\right],
     \label{eq_R2_x3}
\end{align}
where we introduced the length of the screen $\ell_0$, given by
\begin{equation}
   \ell_0\equiv n_1+n_2+1. \label{eq_ell_0}
\end{equation}

Now, using the definitions \equ{eq_phi_definition}-\equ{eq_psi_definition} of $\Phi$ and $\Psi$, the two pairs of equations \equ{eq_pair1_1}-\equ{eq_pair1_2} and \equ{eq_pair2_1}-\equ{eq_pair2_2} can be rewritten as the following $4\times4$ matrix Wiener-Hopf equation:
\begin{equation}
    [\mathbf{U}_{4}]_- + \mathbf{K}_4[\mathbf{U}_{4}]_+ = \mathbf{M}_4^{-1}\mathbf{R}_4,
    \label{eq_matrix_eq_44}
\end{equation}
with known matrix kernel $\mathbf{K}_4=\mathbf{M}_4^{-1}\mathbf{J}_4\mathbf{M}_4$ and known forcing term $\mathbf{R}_4$ defined by
\begin{equation}
\mathbf{M}_4 =\frac{1}{\operatorname{s}(1)} 
\left(
\begin{array}{cccc}
  -\operatorname{s}(\ell_0)  &\operatorname{s}(\ell_0-1) & \operatorname{s}(1)& 0  \\
\operatorname{s}(\ell_1-1) & -\operatorname{s}(\ell_1)  &  0  & 0 \\
 \operatorname{s}(1)  & 0 & -\operatorname{s}(\ell_0) &  \operatorname{s}(\ell_0-1)   \\
  0 & 0  & \operatorname{s}(\ell_2-1) & -\operatorname{s}(\ell_2)  \\
\end{array}
\right), 
\label{eq_M}
\end{equation}
\begin{equation*}
\mathbf{J}_4 = 
\left(
\begin{array}{cccc}
 -1 & 0 & 0 & 0 \\
 0 & 1& 0 & 0 \\
 0 & 0 & -1 & 0 \\
 0 & 0 & 0 & 1\\
\end{array}
\right),
\quad
\mathbf{R}_4=
\left(
\begin{array}{c}
 R_1 \\
 0 \\
R_2 \\
 0 \\
\end{array}
\right),
\end{equation*}
and the unknown `plus' and `minus' vector functions $[\mathbf{U}_{4}]_-$ and $[\mathbf{U}_{4}]_+$ are given by
\begin{equation*}
[\mathbf{U}_{4}]_- =
\left(
\begin{array}{c}
U_-(x,-n_1) \\
U_-(x,-n_1-1)  \\
U_-(x,n_2) \\
U_-(x,n_2+1)\\ 
\end{array}
\right), 
\quad
[\mathbf{U}_{4}]_+=
\left(
\begin{array}{c}
U_+(x,-n_1) \\
U_+(x,-n_1-1)  \\
U_+(x,n_2) \\
U_+(x,n_2+1)\\ 
\end{array}
\right).
\end{equation*}
Here we have expressed the equation in a similar form as in the continuous case in \cite{Erbas2007ScatteringPlates}. Note also that $(\mathbf{J}_4)^2=\mathbf{I}_4$, where $\mathbf{I}_4$ is the unit matrix, meaning that $\mathbf{J}_4$ is involutory. Many interesting matrix Wiener--Hopf problems have a kernel that can be written in this form \cite{Aitken2024OnProblems,Erbas2002ScatteringStructures}.

\subsection{Chebyshev polynomials representation}
\label{sect_cheb}
As previously noted in \cite{Sharma2017OnPolynomials,Mason2002ChebyshevPolynomials}, the function $\operatorname{s}(n),$ defined in \equ{eq_s_def}, can be expressed in terms of Chebyshev polynomials as:
\begin{equation}
\begin{aligned}
    \operatorname{s}(n) &= \operatorname{s}(1)\  \U_{n-1}\left(z\right), & n\geqslant 1,
\end{aligned}    
\label{eq_cheb_def}
\end{equation}
where $\operatorname{s}(1)=(y-y^{-1})$,  $\U_n(z)$ is the $n$-th order Chebyshev polynomial of the second kind\footnote{Note that the standard notation for the Chebyshev polynomials of the second kind is $\text{U}_n(z)$; however, here they are denoted differently for brevity, to avoid confusion with the notation used for discrete Fourier transforms.}, and $z\in[-1,1]$, related to the complex variable $x$ by:
\begin{equation}
    z = \frac{1}{2}(y+y^{-1})
    = -\frac{1}{2}(\Omega^2-4+x+x^{-1}). 
    \label{eq_z_x}
\end{equation}

Thus, \equ{eq_M} may be rewritten as:
\begin{equation}
\mathbf{M}_4 = 
\left(
\begin{array}{cccc}
-\U_{\ell_0-1}(z) &  \U_{\ell_0-2}(z) & 1 & 0 \\
 \U_{\ell_1-2}(z) & -\U_{\ell_1-1}(z) & 0 & 0 \\
 1 & 0 & -\U_{\ell_0-1}(z) &  \U_{\ell_0-2}(z) \\
 0 & 0 &  \U_{\ell_2-2}(z) & -\U_{\ell_2-1}(z) \\
\end{array}
\right),
\label{eq_M_cheb}
\end{equation}
and \equ{eq_R1_x3}-\equ{eq_R2_x3} are rewritten as:
\begin{align}
    R_1(x) &= \Upsilon(x)\left[\operatorname{s}_p(N-\ell_2) - \U_{\ell_0-1}(z) \operatorname{s}_p(\ell_1)\right] + 
    \U_{\ell_0-2}(z)\left[\Pi(x)\operatorname{s}_p(\ell_1-1)
     - u^\text{sc}_{0,-n_1-1}\right],
     \label{eq_R1_x4} \\
     R_2(x) &= \Upsilon(x)\left[ \operatorname{s}_p(\ell_1) - \U_{\ell_0-1}(z)  \operatorname{s}_p(N-\ell_2) \right] +
    \U_{\ell_0-2}(z)
     \left[\Pi(x)\operatorname{s}_p(N-(\ell_2-1))-u^\text{sc}_{0,n_2+1}\right].
     \label{eq_R2_x4}
\end{align}

\begin{remark}
    In the discrete case, every entry in \equ{eq_matrix_eq_44} is expressed through a Chebyshev polynomial of the second kind, which has a finite number of known roots for real $z\in[-1,1]$, and therefore, by \equ{eq_z_x}, only a finite number of poles and zeros in the $x$ complex plane. For the continuous waveguide with Floquet--Bloch periodic boundary conditions on the walls, a matrix equation similar to \equ{eq_matrix_eq_44} is obtained in \cite{Erbas2007ScatteringPlates}. However, in contrast to the discrete case, the continuous analogue of \equ{eq_matrix_eq_44} involves trigonometric terms with an infinite number of poles and zeros.
\end{remark}

\subsection{Symmetric waveguide}
We will now consider in detail the symmetric case corresponding to imposing
\begin{equation}
\ell_1=\ell_2\equiv\ell, \label{eq_symmetry_condition}
\end{equation}
meaning that the gaps on each side of the strip are of equal size.
It can easily be shown from the definitions \equ{eq_mode_form}-\equ{eq_N_def} of $\operatorname{s}_p(\ell)$, $N$ and $\theta$, that if \equ{eq_symmetry_condition} holds, then
\begin{equation}
    \begin{cases}
        \operatorname{s}_p(\ell) = \operatorname{s}_p(N-\ell), & \text{if $p$ is \textit{odd}}, \\
        \operatorname{s}_p(\ell) = -\operatorname{s}_p(N-\ell), & \text{if $p$ is \textit{even}}. 
    \end{cases}    
    \label{equ:simmetric_forcing_condition_on_s}
\end{equation}
We will further discuss the case of odd $p$ in \equ{eq_u_in_duct}, when the incident mode has a maximum in the middle of the cross-section of the waveguide; the even-$p$ case is analogous. It then follows from \equ{equ:simmetric_forcing_condition_on_s} and \equ{eq_R1_x4}-\equ{eq_R2_x4} that:
\begin{align}
    &\Phi(x,-n_1) =  \Phi(x,n_2)\equiv\Phi_0(x), && \Phi(x,-n_1-1) =  \Phi(x,n_2+1)\equiv\Phi_1(x),
    \label{phi_symm}
    \\
    &\Psi(x,-n_1) =  \Psi(x,n_2)\equiv\Psi_0(x), && \Psi(x,-n_1-1) =  \Psi(x,n_2+1)\equiv\Psi_1(x), 
    \label{psi_symm}\\
    &R_1(x) = R_2(x) \equiv R_0(x), &&
    u^\text{sc}_{0,-n_1-1}=u^\text{sc}_{0,n_2+1}\equiv u^*,
     \label{eq_R1_symm}
\end{align}
where, using the condition \equ{equ:simmetric_forcing_condition_on_s} for odd $p$, we find $R_0(x)$ to be:
\begin{equation}
    R_0(x) = \operatorname{s}_p(\ell) \Upsilon(x)\left[1 - \U_{\ell_0-1}(z) \right] + 
    \left[\Pi(x)\operatorname{s}_p(\ell-1)
     - u^*\right]\U_{\ell_0-2}(z).
     \label{eq_R0_symm}
\end{equation}
Moreover, inspired by \equ{eq_phi_definition}-\equ{eq_psi_definition}, and upon introducing the new functions 
\begin{align}
    &U^{(0)}_-(x) := U_-(x,-n_1) =  U_-(x,n_2), &&U^{(1)}_-(x) := U_-(x,-n_1-1) =  U_-(x,n_2+1), 
    \label{eq_U0_def}
    \\
    &U^{(0)}_+(x) := U_+(x,-n_1) =  U_+(x,n_2),     
    &&U^{(1)}_+(x) := U_+(x,-n_1-1) =  U_+(x,n_2+1).
    \label{eq_U1_def}
\end{align}
we can rewrite $\Phi_{0},\ \Phi_{1}$ and $\Psi_{0},\ \Psi_{1}$ as
\begin{align}
    &\Psi_0(x) = U^{(0)}_-(x) - U^{(0)}_+(x),  && \Phi_0(x) = U^{(0)}_-(x) + U^{(0)}_+(x), \label{eq_psi0}\\
    &\Psi_1(x) = U^{(1)}_-(x) - U^{(1)}_+(x), &&\Phi_1(x) = U^{(1)}_-(x) + U^{(1)}_+(x). \label{eq_phi1}
\end{align}

\begin{remark}
Without loss of generality, we will further assume that $n_1=0$; then:
\begin{align}
    &n_1 = 0,
    &&N_1 = \ell, \\
    &n_2 = \ell_0-1, 
    &&N_2 = \ell+\ell_0-1.
\end{align}
Then for $-\ell\leq n<0$ from \equ{eq_ansatz_1A} 
we find: 
\begin{alignat}{2}
    &\Phi(x,n) = \Phi_1(x)\frac{\operatorname{s}(\ell+n)}{\operatorname{s}(\ell-1)}, \label{eq_ansatz_1A3}
\end{alignat}
and for $0\leq n \leq \ell_0-1$ from \equ{eq_ansatz_b}-\equ{eq_R}, \equ{eq_C_expr}-\equ{eq_D_expr}
we find:
\begin{alignat}{2}
    \Psi(x,n) = \left[\Psi_0(x)-\Upsilon(x)\operatorname{s}_p(\ell)\right]\frac{\operatorname{s}(n)+\operatorname{s}(\ell_0-1-n)}{\operatorname{s}(\ell_0-1)} + \Upsilon(x)\operatorname{s}_p(\ell+n).
    \label{eq_ansatz_b_sol}
\end{alignat}
\end{remark}


Thus, using \equ{phi_symm}-\equ{eq_R1_symm}, we can reduce the two pairs of equations \equ{eq_pair1_1}-\equ{eq_pair1_2} and \equ{eq_pair2_1}-\equ{eq_pair2_2} to only two equations:
\begin{alignat}{9}
     [1-\U_{\ell_0-1}(z)] &\Psi_0(x) & &+ 
   & &\U_{\ell_0-2}(z)&&\Psi_1(x) & &= R_0(x), \label{eq_2x2cheb1} 
   \\
   \U_{\ell-2}(z) &\Phi_0(x) & &- 
       & &\U_{\ell-1}(z)&&\Phi_1(x) & &=0, \label{eq_2x2cheb2}
\end{alignat}

Dividing the equation \equ{eq_2x2cheb1} by $(1-\U_{\ell_0-1}(z))$, the equation \equ{eq_2x2cheb2} by $\U_{\ell-1}(z)$ and using \equ{eq_psi0}-\equ{eq_phi1}, we obtain:
\begin{align}
    &\Psi_1(x)\mathcal{K}^{(1)}(x) + \Psi_0(x)   = F(x), \label{eq_symm_pair1_2} \\
    &\Phi_0(x)\mathcal{K}^{(0)}(x) - \Phi_1(x) =0, \label{eq_symm_pair2_2}
\end{align}
where we defined the new functions
\begin{align}
    \mathcal{K}^{(0)}(x)  = \frac{\U_{\ell-2}(z(x))}{\U_{\ell-1}(z(x))}, 
    \qquad
    \mathcal{K}^{(1)}(x)  = \frac{\U_{\ell_0-2}(z(x))}{1-\U_{\ell_0-1}(z(x))},
    \label{eq_K1_1} 
\end{align}
\begin{equation}
    F(x) = \frac{R_0(x)}{1-\U_{\ell_0-1}(z(x))} =  \operatorname{s}_p(\ell) [\Pi(x)+1] + 
    \left[\Pi(x)\operatorname{s}_p(\ell-1)
     - u^*\right]\mathcal{K}^{(1)}(x).
     \label{eq_F_1}
\end{equation}
Finally, after some simple algebraic transformations and using \equ{eq_psi0}-\equ{eq_phi1} we obtain the following $2\times2$ matrix Wiener-Hopf equation
\begin{equation}
    [\mathbf{U}_{2}]_- + \mathbf{K}_2[\mathbf{U}_{2}]_+ = \mathbf{F}_2, \label{eq_matrix_symmetric_sol*}
\end{equation}
where the known matrix kernel $\mathbf{K}_2$ is given\footnote{Similarly to $\mathbf{K}_4$, the matrix kernel $\mathbf{K}_2$ can be written in the form $\mathbf{K}_2=\mathbf{M}_2^{-1}\mathbf{J}_2\mathbf{M}_2$ with $(\mathbf{J}_2)^2=\mathbf{I}_2$, where expressions for $\mathbf{M}_2$ and $\mathbf{J}_2$ can be found from \equ{eq_symm_pair1_2}-\equ{eq_symm_pair2_2}.} by
\begin{equation}
\mathbf{K}_2 = -\frac{1}{1+\K^{(0)}\K^{(1)}} 
\left(
\begin{array}{cccc}
 (1-\K^{(0)}\K^{(1)})
 & 
 2\K^{(1)}
 \\
  2\K^{(0)}
  &  
  -(1-\K^{(0)}\K^{(1)})
  \\
\end{array}
\right),
\label{eq_K2_def}
\end{equation}
and the known forcing vector $\mathbf{F}_2$ is given by
\begin{equation*}
    \mathbf{F}_2 = \frac{1}{1+\K^{(0)}\K^{(1)}}
\left(
\begin{array}{c}
 F \\
 F\K^{(0)} \\
\end{array}
\right).
\end{equation*}
The unknown vector functions $[\mathbf{U}_{2}]_-$ and $[\mathbf{U}_{2}]_+$ are given by
\begin{equation*}
[\mathbf{U}_{2}]_- =
\left(
\begin{array}{c}
U^{(0)}_-  \\
U^{(1)}_- \\
\end{array}
\right),
\quad
[\mathbf{U}_{2}]_+=
\left(
\begin{array}{c}
U^{(0)}_+  \\
U^{(1)}_+ \\
\end{array}
\right).
\end{equation*}

Just as in the general case (see \sect{sect_cheb}), the matrix \equ{eq_K2_def} is meromorphic with a finite number of poles and zeros, but now it is a $2\times2$ system.
Note that the equation \equ{eq_matrix_symmetric_sol*} is valid in some annulus containing the unit circle when $\Omega$ has a small imaginary part, where all the terms in the equation are analytic.

\section{On the matrix kernel factorisation for the symmetric case}
\label{sect_matrix_fact}
The crucial step in the Wiener--Hopf method is the factorisation of the kernel \equ{eq_K2_def}, such that $\mathbf{K}_2$ is the product of $[\mathbf{K}_2]_-$ and $[\mathbf{K}_2]_+,$ analytic inside and outside the unit circle, respectively. However, whereas factorisation of a scalar kernel is relatively straightforward \cite{NobleB1958MethodsPDEs}, a rigorous factorisation of a matrix kernel is substantially more difficult \cite{Kisil2021TheMethods, Rogosin2016ConstructiveMatrix-functions}, with no known general procedure available. In this section, we will discuss several approaches that may potentially be applied for factorising the matrix kernel \equ{eq_K2_def}.

\subsection{Rational matrix factorisation}
The structure of the matrix \equ{eq_K2_def} implies that certain known methods for factorising matrices with polynomial entries may be applied here. In particular, an explicit algorithm for the Wiener--Hopf factorisation has been developed by Adukov et al.~\cite{Adukov2022AnPackage}, based on the method introduced in \cite{Adukov1999FactorizationFunctions}. This technique is designed for the exact factorisation of \(2 \times 2\) matrix functions with Laurent polynomial entries\footnote{Laurent polynomials differ from ordinary polynomials in that they may contain terms of negative degree.} and rational coefficients, thereby avoiding the numerical instabilities possible in approximate approaches.
The authors implemented this approach in the \texttt{ExactMPF} package for \textsc{Maple} \cite{Adukov2022AnPackage}, which enables the exact decomposition of matrix polynomials through symbolic computation.
Conditions for the stability of the factorisation are discussed in \cite{Adukov2020ExactFunction, Adukova2024AnPackage}. 

Within this work, the applicability of the \texttt{ExactMPF} package has been investigated for the matrix \equ{eq_K2_def}.
We explored two possible approaches to applying this algorithm.

Our first approach is to premultiply the matrix \equ{eq_K2_def} by its common denominator to obtain a polynomial matrix, which can then be supplied directly to the \texttt{ExactMPF} package. This method is applicable only when the determinant is exactly factorisable, that is, expressible as a product of polynomials whose roots lie either strictly inside or strictly outside the unit circle. For the present matrix, this condition holds only when the imaginary part of $\Omega$ is positive, $\varepsilon_\Omega>0$. Even then, the method breaks down at the determinant factorisation stage: the high-order polynomials that arise, even in the simplest admissible case $\ell_0=2$ and $\ell=2$, exceed the limits of exact arithmetic, preventing the algorithm from reliably determining whether their roots lie inside or outside the unit circle.

For the second approach, we may notice that the determinant of the original matrix \equ{eq_K2_def} is always equal to $1$, while its entries are rational functions. To apply the \texttt{ExactMPF} package, these entries must be approximated by Laurent polynomials on the unit circle for $\varepsilon_\Omega > 0$. However, once such an approximation is introduced, the determinant is no longer equal to $1$. 
Alternatively, we may transform the matrix \equ{eq_K2_def} into the form described in \cite{Adukov2020ExactFunction}, which guarantees that the determinant of the resulting matrix remains constant and equal to $1$, even when the entries are approximated by Laurent polynomials. The difficulty with this approach is that the smaller the imaginary part of $\Omega$, $\varepsilon_\Omega$, the less smooth the matrix entries become on the unit circle. Consequently, for any physically meaningful $\varepsilon_\Omega$, obtaining a sufficiently accurate approximation requires Laurent polynomials of extremely high order, which are not compatible with the algorithm.

Finally, we may conclude that, although the algorithm, developed in \cite{Adukov2022AnPackage}, is in principle applicable to the factorisation of the matrix \equ{eq_K2_def}, its practical implementation becomes complex and will not be pursued in this work.


\subsection{Khrapkov-Daniele type of the symmetric matrix kernel in a special case}
There is another approach to the explicit factorisation of the matrix \equ{eq_K2_def}, provided there is an additional restriction on the screen length $\ell_0$ and the gap length $\ell$. 
In \cite{Erbas2007ScatteringPlates} it was shown that, in the particular case where the barrier length is twice the gap length, the continuous analogue of $\mathbf{K}_2$ \equ{eq_K2_def} belongs to the Khrapkov--Daniele class introduced in \cite{Khrapkov1971CertainForces,Daniele1978OnMethod}. This class of matrices admits \textit{commutative factorisation} \cite{Shanin2005NecessaryMatrices} and can therefore always be reduced to scalar factorisation problems. Square matrices of arbitrary size possessing the Khrapkov--Daniele structure can be written in the form
\begin{equation}
    \mathbf{K}(x) = g_1(x)\mathbf{I} + g_2(x)\mathbf{J}(x), \label{eq:Krapkov_intro}
\end{equation}
where $\mathbf{I}$ is the identity matrix, $g_1(x),g_2(x)$ are arbitrary scalar functions of $x$ with algebraic behaviour at infinity, and $\mathbf{J}(x)$ is a square matrix with entire elements of algebraic form, such that
\begin{equation}
    \mathbf{J}^2(x) = \Delta(x)\mathbf{I},
    \label{eq_delta_J2_cond}
\end{equation}
where $\Delta(x)$ is a polynomial in $x$. When $\Delta(x)=O(|x|^\alpha), x\to\infty$ where $\alpha\leqslant2$, the factorisation can be constructed explicitly as described in \cite{Veitch2007OnEigenvalues}.

The matrix \equ{eq_K2_def} does not immediately have the structure of \equ{eq:Krapkov_intro}.
However, we note that, similarly to \equ{eq_cheb_def}, we may express the  $n$-th order Chebyshev polynomial of the first kind \cite{Mason2002ChebyshevPolynomials, Sharma2017OnPolynomials} as
$$
\T_n\left(z\right)=\frac{1}{2}(y^n+y^{-n}),
$$
where $z(x)$ is defined by \equ{eq_z_x}. Then, introducing the matrix functions
\begin{equation}
\mathbf{J}_2 =  
\left(
\begin{array}{cccc}
 -1 & 0 \\
  0 &  1 \\
\end{array}
\right), 
\qquad
\mathbf{S} =  
\left(
\begin{array}{cccc}
 1 & 0 \\
  \T_1(z) &  1 \\
\end{array}
\right),
\label{eq_S_transf}
\end{equation}
we can show that
\begin{equation}
    \mathbf{S}^{-1}\mathbf{J}_2\mathbf{K}_2\mathbf{S}   = 
    \frac{1}{\T_{\ell+\ell_0'}(z)}
    \left(
    \begin{array}{cccc}
-\T_{\ell-\ell_0'}(z) &
-2\T_{\ell_0'}(z)\U_{\ell-1}(z)
\\
2(\gamma/2)^2 \T_{\ell}(z)\U_{\ell_0'-1}(z) &
-\T_{\ell-\ell_0'}(z) ,
  \\
\end{array}
\right),
\label{eq_transf_M}
\end{equation}
with
\begin{equation}
    \gamma=\operatorname{s}(1)=\sqrt{\Lambda(x)^2-4},
\qquad
    \ell_0'=\frac{\ell_0-1}{2},
\end{equation}
where $\ell_0'$ must be an integer parameter, equal to half the length of the screen for odd values of $\ell_0$.

It is straightforward to show that in the special case $\ell=\ell_0'$ the matrix function \equ{eq_transf_M} takes the form
\begin{equation}
\mathbf{S}^{-1}\mathbf{J}_2\mathbf{K}_2\mathbf{S}
=g_1(x)\mathbf{I} + g_2(x)\mathbf{J}(x),
\label{eq:Kh_factor_intro}
\end{equation}
where
\begin{equation}
    g_1(x)=-\frac{1}{\T_{2\ell}\left(z(x)\right)}, \quad g_2(x)=\frac{\T_{\ell}\left(z(x)\right)\U_{\ell-1}\left(z(x)\right)}{x\T_{2\ell}\left(z(x)\right)},
    \quad
    \mathbf{J}=\left(
    \begin{array}{cccc}
     0 & -2x \\
     \tfrac{1}{2}x\gamma^2 & 0 \\
    \end{array}
    \right).
\end{equation}
The matrix function $\mathbf{J}(x)$ satisfies the condition \equ{eq_delta_J2_cond} with
\begin{equation}
    \Delta(x)=-x^2(\gamma(x))^2 = -\left[x^2+(\Omega^2-6)x+1\right]\left[x^2+(\Omega^2-2)x+1\right],
\end{equation}
which is a polynomial of degree $4$. Therefore, we have shown that, in the discrete case also, we have a matrix of the Khrapkov--Daniele class. However, when the degree of $\Delta(x)$ exceeds $2$, the factors in \equ{eq:Kh_factor_intro} exhibit exponential growth at infinity \cite{Abrahams1998OnType}, which prevents the direct application of Liouville's theorem. Several approaches to overcome this difficulty have been discussed in the literature \cite{Antipov2002FactorizationTheory, Daniele1984OnProblems, Abrahams1998OnType}. These procedures are all quite complex and lie beyond the scope of this work. 
Note that in the continuous case this problem does not arise, as the corresponding polynomial degree is less than 2 \cite{Erbas2007ScatteringPlates}, and hence the factorisation is easier.

Importantly, even if we manage to factorise the matrix \equ{eq_transf_M}, it will not allow us to solve the Wiener--Hopf equation \equ{eq_matrix_symmetric_sol*} straight away, as the matrix function $\mathbf{S}$ has two reciprocal poles, as follows from its structure. Thus, an additional step, i.e., the application of the pole removal method \cite{Lawrie1993AcousticEdges,Lawrie1994AcousticDucts} is required. Alternatively, we may follow the procedure of the direct continuous-discrete analogy, described in \cite{Korolkov2025OnProblems} (see remark below).

\begin{remark}
    To derive the Wiener--Hopf equation with the kernel in the standard Khrapkov--Daniele form, we may transform the unknowns in \equ{eq_matrix_symmetric_sol*} as follows:
\begin{equation}
    [\mathbf{U}_{2}]_{\pm} = \mathbf{S}\mathbf{W}_{\pm}\mp\mathbf{T}, \label{eq_link_unknowns}
\end{equation}
where $\mathbf{S}$ is defined by \equ{eq_S_transf} and $\mathbf{T}$ is a known column function. This substitution introduces new unknown functions $\mathbf{W}_{\pm}$, which coincide with the unknowns in the Wiener--Hopf equation formulated for the symmetric problem derived from the second Green’s identity, as described in \cite{Korolkov2025OnProblems}. Further details are omitted for brevity. Substitution of \equ{eq_link_unknowns} into \equ{eq_matrix_symmetric_sol*} yields the matrix Wiener--Hopf equation with the kernel in the form $\mathbf{S}^{-1}\mathbf{K}_2\mathbf{S}$.
\end{remark}

\subsection{From kernel factorisation to additive splitting}
Although, based on the above, we will not pursue the direct factorisation of the matrix \equ{eq_K2_def}, its rational structure implies that we can circumvent this step by applying the pole removal technique, a standard method frequently used in Wiener--Hopf problems (see \cite{Lawrie1993AcousticEdges,Lawrie1994AcousticDucts,Aitken2024OnProblems}). The method yields an algorithmic procedure for additive splitting of a rational function, in which the pole contributions are subtracted via the residue theorem.

This method has also been employed in \cite{Erbas2007ScatteringPlates, Aitken2024OnProblems}; in our case, however, the application of the method is simpler and more straightforward, since the matrix \equ{eq_K2_def} has only a finite number of poles. This enables an exact solution, in contrast to the case of the continuous waveguide. The procedure is discussed in the next section.

%% file: 4_pole_removal.tex
\section{Solution via pole removal for the symmetric case}
\label{sect_pole_removal}
\subsection{Polynomial representation of matrix terms}
Since the zeros and poles of \equ{eq_K1_1} are determined from the roots of the Chebyshev polynomials 
(see \appx{sect:cheb_roots} for details), we can represent $\mathcal{K}^{(0)}(x)$ and $\mathcal{K}^{(1)}(x)$ as rational functions:
\begin{align}
    \mathcal{K}^{(0)}(x) & = -x\, \frac{P^{(0)}(x)}{Q^{(0)}(x)}, \qquad 
    \mathcal{K}^{(1)}(x) = x\, \frac{P^{(1)}(x)}{Q^{(1)}(x)}, \label{eq_def_Ks}
\end{align}
where, for $s\in\{0,1\}$, we define the polynomials
\begin{align}
P^{(s)}(x) &= \prod_{j=1}^{J^{(s)}-1} \left(x-\mu^{(s)}_{j,+}\right)\left(x-\mu^{(s)}_{j,-}\right), \label{eq_P}\\
Q^{(s)}(x) &= \prod_{j=1}^{J^{(s)}} \left(x-\lambda^{(s)}_{j,+}\right)\left(x-\lambda^{(s)}_{j,-}\right), \label{eq_Q}
\end{align}
with $J^{(s)}$ given by
\begin{align*}
    J^{(1)} = \left\lfloor \tfrac{\ell_0}{2} \right\rfloor, \qquad
    J^{(0)} = \ell-1.
\end{align*}
The zeros $\mu^{(s)}_{j,\pm}$ and poles $\lambda^{(s)}_{j,\pm}$ of \equ{eq_def_Ks}
are determined in \appx{sect:cheb_roots}, namely \equ{eq_mu0_def}-\equ{eq_lambda0_def}, \equ{eq_mu1_def}-\equ{eq_lambda1_def}.
Note that all poles and zeros are reciprocal, that is,
\begin{align}
    \mu^{(s)}_{j,+}\,\mu^{(s)}_{j,-} = 1, \qquad 
    \lambda^{(s)}_{j,+}\,\lambda^{(s)}_{j,-} = 1. \label{eq_poles_reciprocity}
\end{align}

\subsection{Additive splitting via pole removal}
\label{sect_add_split}
Since both $\mathcal{K}^{(0)}(x), \ \mathcal{K}^{(1)}(x)$ in \equ{eq_K1_1} are rational, they contain only a finite number of poles. We can therefore additively split the terms in the system \equ{eq_symm_pair1_2}-\equ{eq_symm_pair2_2} by use of the pole removal technique \cite{Lawrie1993AcousticEdges,Lawrie1994AcousticDucts}.

Thus, we rewrite the equations \equ{eq_symm_pair1_2}-\equ{eq_symm_pair2_2} as
\begin{alignat}{7}
    &\left[U^{(1)}_-\mathcal{K}^{(1)}\right]_-  -\left[U^{(1)}_+\mathcal{K}^{(1)}\right]_- & &+& & U^{(0)}_- - F_-  & &= U^{(0)}_+ - \left[U^{(1)}_-\mathcal{K}^{(1)}\right]_+ + \left[U^{(1)}_+\mathcal{K}^{(1)}\right]_+ + F_+, \label{eq_symm_pair2_22}
    \\
    &\left[U^{(0)}_-\mathcal{K}^{(0)}\right]_- + \left[U^{(0)}_+\mathcal{K}^{(0)}\right]_- & &-& &U^{(1)}_- & &= U^{(1)}_+ - \left[U^{(0)}_-\mathcal{K}^{(0)}\right]_+ - \left[U^{(0)}_+\mathcal{K}^{(0)}\right]_+, \label{eq_symm_pair1_22}
\end{alignat}
where the additively split terms are defined by
\begin{align}
     \left[U^{(s)}_-\mathcal{K}^{(s)}\right]_- &= U^{(s)}_-\mathcal{K}^{(s)} - \left[U^{(s)}_-\mathcal{K}^{(s)}\right]_+,
     \label{eq_split_term_neg}
     \\
     \left[U^{(s)}_+\mathcal{K}^{(s)}\right]_+ &= U^{(s)}_+\mathcal{K}^{(s)} - \left[U^{(s)}_+\mathcal{K}^{(s)}\right]_-, \label{eq_split_term_pos}
     \\
     \left[U^{(s)}_-\mathcal{K}^{(s)}\right]_+ &= \sum_{j=1}^{J^{(s)}}  w^{(s)}_{j,-}
     b^{(s)}_{j,-}\frac{\lambda^{(s)}_{j,-}}{x-\lambda^{(s)}_{j,-}}, \label{eq_split_term_mp}
    \\
    \left[U^{(s)}_+\mathcal{K}^{(s)}\right]_- 
    &= \sum_{j=1}^{J^{(s)}}  
     w^{(s)}_{j,+}
     b^{(s)}_{j,+}\frac{\lambda^{(s)}_{j,+}}{x-\lambda^{(s)}_{j,+}}. \label{eq_split_term_pm}
\end{align}
In \equ{eq_split_term_neg}-\equ{eq_split_term_pm} the constants $w^{(s)}_{j,\pm}$, $b^{(s)}_{j,\pm}$ are found from the residue theorem:
\begin{align}
    w^{(0)}_{j,\pm} = U^{(0)}_\pm\left(\lambda^{(0)}_{j,\pm}\right), 
    \qquad 
    w^{(1)}_{j,\pm} = U^{(1)}_\pm\left(\lambda^{(1)}_{j,\pm}\right),
    \label{eq_w_pm_lambda}
    \\
    b^{(0)}_{j,\pm} = -\frac{P^{(0)}\left(\lambda^{(0)}_{j,\pm}\right)}{Q^{(0)\pm}_j\left(\lambda^{(0)}_{j,\pm}\right)}, \qquad b^{(1)}_{j,\pm} = \frac{P^{(1)}\left(\lambda^{(1)}_{j,\pm}\right)}{Q^{(1)\pm}_j\left(\lambda^{(1)}_{j,\pm}\right)},        
\end{align}
where
\begin{align}
    Q^{(s)}_{j,\pm} = \frac{Q^{(s)}(x)}{x-\lambda^{(s)}_{j,\pm}} =\prod_{i=1}^{J^{(s)}}\left(x-\lambda^{(s)}_{i,\mp}\right) \prod_{\substack{i=1\\i\neq j}}^{J^{(s)}}\left(x-\lambda^{(s)}_{i,\pm}\right). 
    \label{eq_Q_j}
\end{align}

Similarly, we split $F(x)=F_-(x)+F_+(x)$ in \equ{eq_F_1} as:
\begin{alignat}{9}
    F_-(x) &=& &
    \operatorname{s}_p(\ell) \Pi_-(x) &
    &+&
    \operatorname{s}_p(\ell-1)\left[\Pi(x)\mathcal{K}^{(1)}(x)\right]_-
    - 
    u^*\left[\mathcal{K}^{(1)}(x)\right]_-, \label{eq_F_minus}\\
    F_+(x) &=& &
    \operatorname{s}_p(\ell) [1+\Pi_+(x)] &
    &+&
    \operatorname{s}_p(\ell-1) \left[\Pi(x)\mathcal{K}^{(1)}(x)\right]_+ 
    -
    u^*\left[\mathcal{K}^{(1)}(x)\right]_+ \label{eq_F_plus},
\end{alignat}
where the terms in \equ{eq_F_minus}-\equ{eq_F_plus} are defined as:
\begin{align}
    \Pi_-(x) &= \frac{x}{x-(x_p)^{-1}},
    \label{eq_ref_limits1}
    \\
    1+\Pi_+(x) &= 1+\frac{x_p}{x-x_p} = \frac{x}{x-x_p},
    \\
    \left[\mathcal{K}^{(1)}(x)\right]_+ &= \sum_{j=1}^{J^{(1)}}b^{(1)}_{j,-}\frac{\lambda^{(1)}_{j,-}}{x-\lambda^{(1)}_{j,-}} 
    \\
    \left[\mathcal{K}^{(1)}(x)\right]_- &= \mathcal{K}^{(1)}(x) - \left[\mathcal{K}^{(1)}(x)\right]_+,
    \\
    \left[\Pi(x)\mathcal{K}^{(1)}(x)\right]_+ &= \mathcal{K}^{(1)}(x_p)\frac{x_p}{x-x_p} + \sum_{j=1}^{J^{(1)}}\Pi\left(\lambda^{(1)}_{j,-}\right)b^{(1)}_{j,-}\frac{\lambda^{(1)}_{j,-}}{x-\lambda^{(1)}_{j,-}},
    \\
    \left[\Pi(x)\mathcal{K}^{(1)}(x)\right]_- &= \Pi(x)\mathcal{K}^{(1)}(x) - \left[\Pi(x)\mathcal{K}^{(1)}(x)\right]_+.
    \label{eq_ref_limits2}
\end{align}

\subsection{Application of Liouville's theorem}
Next, we analyse the behaviour of the functions in \equ{eq_symm_pair2_22}-\equ{eq_symm_pair1_22} at $0$ and at $\infty$. In particular, we find the limits of \equ{eq_F_minus}-\equ{eq_F_plus} as $x \to 0$ and $x \to \infty$, respectively (see \appx{appx_limits} for details):
\begin{align}
     \lim_{x\to0}F_-&=\operatorname{s}_p(\ell-1)f_p - \displaystyle u^*f_*, 
     \\
     \lim_{x\to\infty}F_+ &=\operatorname{s}_p(\ell),
\end{align}
where the known constants $f_p$ and $f_*$ are defined by
\begin{align}
    f_p=\displaystyle \mathcal{K}^{(1)}(x_p) +\sum_{j=1}^{J^{(1)}}\Pi\left(\lambda^{(1)}_{j,-}\right)b^{(1)}_{j,-},
    \qquad
    f_*= \sum_{j=1}^{J^{(1)}}b^{(1)}_{j,-}. \label{eq_f_star}
\end{align}
\FloatBarrier

Note that all the terms in \equ{eq_symm_pair2_22}-\equ{eq_symm_pair1_22} remain bounded at infinity, and both the left-hand side and the right-hand side tend to constants as $x \to 0$ and $x \to \infty$, respectively. Therefore, by analytic continuation, we conclude that the left-hand side and the right-hand side of each equation are equal to the same entire function, which is, by Liouville's theorem, a constant. We denote the constants by $C_1$ for \equ{eq_symm_pair2_22} and $C_0$ for \equ{eq_symm_pair1_22}, such that:
\begin{align}
C_1 &= \sum_{j=1}^{J^{(1)}}  w^{(1)}_{j,-}b^{(1)}_{j,-} + \sum_{j=1}^{J^{(1)}}  w^{(1)}_{j,+}b^{(1)}_{j,+} - \operatorname{s}_p(\ell-1)f_p + u^*f_*=   -\operatorname{s}_p(\ell) + \operatorname{s}_p(\ell)=0\label{eq_C1_fin}\\
    C_0 
    &=\sum_{j=1}^{J^{(0)}}  w^{(0)}_{j,-}b^{(0)}_{j,-} - \sum_{j=1}^{J^{(0)}}  w^{(0)}_{j,+}b^{(0)}_{j,+} = u^*  
    \label{eq_C0_fin}
\end{align}

Note also, that by the symmetry of the problem \equ{eq_U_symmetry}, the reciprocity of the poles \equ{eq_poles_reciprocity}, and the boundary conditions \equ{eq_BC_symmetric1}--\equ{eq_BC_symmetric2}, we have:
\begin{align}
    w^{(0)}_{j,-} &= w^{(0)}_{j,+} + \operatorname{s}_p(\ell), \label{eq_const_sym1}\\
    w^{(1)}_{j,-} &= w^{(1)}_{j,+} - u^*. \label{eq_const_sym2}
\end{align}

Using \equ{eq_const_sym1}-\equ{eq_const_sym2},  we can find the unknown $u^*$ from \equ{eq_C0_fin} as:
\begin{align}
    u^* &= \sum_{j=1}^{J^{(0)}}  w^{(0)}_{j,+} \left[b^{(0)}_{j,-} - b^{(0)}_{j,+}\right]+ \operatorname{s}_p(\ell)\sum_{j=1}^{J^{(0)}}b^{(0)}_{j,-}. \label{eq_unknown_1}
\end{align}

Then, by equating the right-hand sides of \equ{eq_symm_pair2_22}-\equ{eq_symm_pair1_22} with \equ{eq_C1_fin}-\equ{eq_C0_fin}, respectively, and using \equ{eq_split_term_neg}-\equ{eq_split_term_pos}, we obtain:
\begin{alignat}{4}
    U^{(0)}_+ &=&  - &U^{(1)}_+\mathcal{K}^{(1)}  
    + \left(\left[U^{(1)}_-\mathcal{K}^{(1)}\right]_+ + \left[U^{(1)}_+\mathcal{K}^{(1)}\right]_-
     \right) 
    - F_+. \label{eq_symm_pair2_4} \\
    U^{(1)}_+ &=& &U^{(0)}_+\mathcal{K}^{(0)} + \left( \left[U^{(0)}_-\mathcal{K}^{(0)}\right]_+ - \left[U^{(0)}_+\mathcal{K}^{(0)}\right]_-  \right) + C_0.
    \label{eq_symm_pair1_4}
\end{alignat}

From \equ{eq_const_sym1}-\equ{eq_const_sym2} we find
\begin{align}
    \left[U^{(1)}_-\mathcal{K}^{(1)}\right]_+ &= \sum_{j=1}^{J^{(1)}}  w^{(1)}_{j,+} b^{(1)}_{j,-}\frac{\lambda^{(1)}_{j,-}}{x-\lambda^{(1)}_{j,-}} - \mathcal{C}^{(1)}(x)u^*, \label{eq_term_symmetry2}\\
    \left[U^{(0)}_-\mathcal{K}^{(0)}\right]_+ &= \sum_{j=1}^{J^{(0)}}  w^{(0)}_{j,+} b^{(0)}_{j,-}\frac{\lambda^{(0)}_{j,-}}{x-\lambda^{(0)}_{j,-}} + \mathcal{C}^{(0)}(x)\operatorname{s}_p(\ell), \label{eq_term_symmetry1}
\end{align}
where we introduced the functions
\begin{align}
    \mathcal{C}^{(1)}(x) = \sum_{j=1}^{J^{(1)}}  b^{(1)}_{j,-}\frac{\lambda^{(1)}_{j,-}}{x-\lambda^{(1)}_{j,-}}, \qquad 
    \mathcal{C}^{(0)}(x)  = \sum_{j=1}^{J^{(0)}}  b^{(0)}_{j,-}\frac{\lambda^{(0)}_{j,-}}{x-\lambda^{(0)}_{j,-}}. \label{eq_C1_func}
\end{align}

Using the definitions \equ{eq_split_term_mp}-\equ{eq_split_term_pm} and \equ{eq_term_symmetry2}-\equ{eq_term_symmetry1} we hence find:
\begin{align}
\left[U^{(1)}_-\mathcal{K}^{(1)}\right]_+ + \left[U^{(1)}_+\mathcal{K}^{(1)}\right]_- 
    & = 
    \sum_{j=1}^{J^{(1)}}  w^{(1)}_{j,+} \left[ b^{(1)}_{j,-}\frac{\lambda^{(1)}_{j,-}}{x-\lambda^{(1)}_{j,-}} + b^{(1)}_{j,+}\frac{\lambda^{(1)}_{j,+}}{x-\lambda^{(1)}_{j,+}}\right] - u^*\mathcal{C}^{(1)}, \label{eq_split_diff2}\\
    \left[U^{(0)}_-\mathcal{K}^{(0)}\right]_+ - \left[U^{(0)}_+\mathcal{K}^{(0)}\right]_-  
    & = 
    \sum_{j=1}^{J^{(0)}}  w^{(0)}_{j,+} \left[ b^{(0)}_{j,-}\frac{\lambda^{(0)}_{j,-}}{x-\lambda^{(0)}_{j,-}} - b^{(0)}_{j,+}\frac{\lambda^{(0)}_{j,+}}{x-\lambda^{(0)}_{j,+}} \right] + \operatorname{s}_p(\ell)\mathcal{C}^{(0)}. \label{eq_split_diff1}
\end{align}
Finally, noting that $\left[\mathcal{K}^{(1)}\right]_+ = \mathcal{C}^{(1)}$ in the definition of $F_+$ \equ{eq_F_plus}, and combining \equ{eq_C1_fin}-\equ{eq_C0_fin}, \equ{eq_symm_pair2_4}-\equ{eq_symm_pair1_4}, and \equ{eq_split_diff2}-\equ{eq_split_diff1}, we arrive at the following system of equations:
\begin{alignat}{4}
    U^{(0)}_+ &=&  - &U^{(1)}_+\mathcal{K}^{(1)}  
    +
    \sum_{j=1}^{J^{(1)}}  w^{(1)}_{j,+}\mathcal{F}^{(1)}_j
    + \mathcal{G}_1 . \label{eq_symm_pair2_5}
    \\
    U^{(1)}_+ &=& &U^{(0)}_+\mathcal{K}^{(0)} +  \sum_{j=1}^{J^{(0)}}  w^{(0)}_{j,+}\mathcal{F}^{(0)}_j
    + \mathcal{G}_0, \label{eq_symm_pair1_5}
\end{alignat}
where we introduced the following new, explicitly known functions
\begin{align}
    \mathcal{F}^{(1)}_j(x) &= b^{(1)}_{j,-}\frac{\lambda^{(1)}_{j,-}}{x-\lambda^{(1)}_{j,-}} + b^{(1)}_{j,+}\frac{\lambda^{(1)}_{j,+}}{x-\lambda^{(1)}_{j,+}},
    \label{eq_F1}
    \\
    \mathcal{F}^{(0)}_j(x) &= b^{(0)}_{j,-}\frac{x}{x-\lambda^{(0)}_{j,-}} - b^{(0)}_{j,+}\frac{x}{x-\lambda^{(0)}_{j,+}},
    \\
    \mathcal{G}_0(x) &= \operatorname{s}_p(\ell)\sum_{j=1}^{J^{(0)}}b^{(0)}_{j,-} \frac{x}{x-\lambda^{(0)}_{j,-}}.
    \label{eq_G0}
    \\
    \mathcal{G}_1(x) &= 
    -\mathcal{P}(x) - \operatorname{s}_p(\ell-1)
    \sum_{j=1}^{J^{(1)}}\Pi\left(\lambda^{(1)}_{j,-}\right) b^{(1)}_{j,-} \frac{\lambda^{(1)}_{j,-}}{x-\lambda^{(1)}_{j,-}},
    \\
    \mathcal{P}(x) &= \frac{x\operatorname{s}_p(\ell) + x_p\operatorname{s}_p(\ell-1)\mathcal{K}^{(1)}(x_p)}{x-x_p}.
    \label{eq_Pxp}
\end{align}
Note that using \equ{eq_cheb_def} and \equ{eq_K1_1}, we can rewrite $\mathcal{K}^{(1)}$ as
\begin{equation}
\mathcal{K}^{(1)}(x) = \frac{\operatorname{s}(\ell_0-1)}{\operatorname{s}(1)-\operatorname{s}(\ell_0)}, 
\quad \text{hence} \quad
\mathcal{K}^{(1)}(x_p) = \frac{\operatorname{s}_p(\ell_0-1)}{\operatorname{s}_p(1)-\operatorname{s}_p(\ell_0)},
\label{eq_K_sp}
\end{equation}
and that by definition of $\operatorname{s}(n)$ \equ{eq_s_def} for any integer $\ell, \ell_0\geqslant2$ :
\begin{equation}
     \operatorname{s}(\ell-1)\operatorname{s}(\ell_0-1)= \operatorname{s}(\ell)\operatorname{s}(\ell_0)-\operatorname{s}(1)\operatorname{s}(\ell).  \label{eq_l_l0_property2}
\end{equation}
Substituting \equ{eq_K_sp}-\equ{eq_l_l0_property2} into \equ{eq_Pxp} allows one to show that the function $\mathcal{P}(x)$ has a removable singularity at $x = x_p$ and therefore is analytic everywhere.



\subsection{Linear system to determine the constants}
\label{sect_lenar_system}
Substituting \equ{eq_symm_pair2_5} into \equ{eq_symm_pair1_5}, we obtain
\begin{alignat}{4}
    U^{(1)}_+ &=& & \displaystyle \frac{ \mathcal{K}^{(0)}\sum_{j=1}^{J^{(1)}}  w^{(1)}_{j,+}\mathcal{F}^{(1)}_j 
    +  \sum_{j=1}^{J^{(0)}}  w^{(0)}_{j,+}\mathcal{F}^{(0)}_j
    + \left[\mathcal{G}_0 + \mathcal{K}^{(0)}\mathcal{G}_1\right]}{1+\mathcal{K}^{(0)}\mathcal{K}^{(1)}}. \label{eq_symm_pair1_6}
\end{alignat}

Equations \equ{eq_symm_pair1_6} and \equ{eq_symm_pair2_5} determine $U^{(1)}_+$ and $U^{(0)}_+$ up to a finite set of $J^{(0)}+J^{(1)}$ unknown coefficients $w^{(1)}_{j,+}$ ($j=1,\dots,J^{(1)}$) and $w^{(0)}_{j,+}$ ($j=1,\dots,J^{(0)}$), defined by \equ{eq_w_pm_lambda}. These coefficients are precisely the values of $U^{(1)}_+$ and $U^{(0)}_+$ at the corresponding sets of poles.

To proceed, we analyse the poles and zeros of the denominator in \equ{eq_symm_pair1_6}. 
Using \equ{eq_K1_1}, the denominator can be expressed in terms of Chebyshev polynomials as
\begin{equation}
    \frac{1}{1+\mathcal{K}^{(0)}\mathcal{K}^{(1)}} 
    = \frac{(1-\U_{\ell_0-1})\U_{\ell-1}}{(1-\U_{\ell_0-1})\U_{\ell-1} + \U_{\ell_0-2}\U_{\ell-2}}
    = \frac{Q^{(0)}(x)Q^{(1)}(x)}{W(x)}, \label{eq_system_sol2}
\end{equation}
where $Q^{(0)}$ and $Q^{(1)}$ are defined in \equ{eq_Q}, and
\begin{equation}
   W(x) = \prod_{j=1}^{J} \left(x-\nu^{+}_j\right)\left(x-\nu^{-}_j\right), \label{eq_V} 
\end{equation}
with $\nu^{\pm}_j$ determined from the roots of the 
Chebyshev polynomial combination by \equ{eq_nupm_def}, as detailed in \appx{sect:cheb_roots}. Note that some of these solutions coincide with the zeros of \equ{eq_system_sol2}, which can be determined as in \appx{sect:cheb_roots}.

All the poles in \equ{eq_symm_pair1_6} are cancelled by the zeros of \equ{eq_system_sol2}. Moreover, \equ{eq_system_sol2} has $J$ poles $\nu_i^+$ in the $+$ region, whereas $U_+^{(1)}$ is, by definition, analytic in this region. To satisfy this requirement, the numerator of \equ{eq_symm_pair1_6} must therefore vanish at each $\nu_i^+$, $i \in \{1,\dots,J\}$, so that
\begin{equation}
    \mathcal{K}^{(0)}\left(\nu_i^+\right)\sum_{j=1}^{J^{(1)}}  w^{(1)}_{j,+}\mathcal{F}^{(1)}_j\left(\nu_i^+\right) +  \sum_{j=1}^{J^{(0)}}  w^{(0)}_{j,+}\mathcal{F}^{(0)}_j\left(\nu_i^+\right)
    + \mathcal{P}_i=0,\quad i\in\{1,J\}, \label{eq_algebraic_system_coeff}
\end{equation}
where
$$\mathcal{P}_i=\mathcal{G}_0\left(\nu_i^+\right) + \mathcal{K}^{(0)}\left(\nu_i^+\right)\mathcal{G}_1\left(\nu_i^+\right).$$

If $J = J^{(0)}+J^{(1)}$, then \equ{eq_algebraic_system_coeff} is a complete linear system of equations, from which the values of $w^{(1)}_{j,+}$ ($j=1,\dots,J^{(1)}$) and $w^{(0)}_{j,+}$ ($j=1,\dots,J^{(0)}$) can be determined.

If in \equ{eq_V} $J<J^{(0)}+J^{(1)}$, then the system \equ{eq_algebraic_system_coeff} is not complete. Such a situation may occur when, in \equ{eq_system_sol2}, some poles of $\mathcal{K}^{(0)}$ or $\mathcal{K}^{(1)}$ are cancelled by zeros of $\mathcal{K}^{(1)}$ or $\mathcal{K}^{(0)}$, respectively. In this case, if a pole $\lambda_q^{(1)\pm}$ is cancelled, we evaluate \equ{eq_symm_pair1_6} at this point, thereby adding the missing equations to \equ{eq_algebraic_system_coeff}. If instead a pole $\lambda_q^{(0)\pm}$ is cancelled, we consider \equ{eq_symm_pair1_6} multiplied by $(x-\lambda_q^{(0)\pm})$ and then evaluate it at this pole. Therefore, \equ{eq_algebraic_system_coeff} is always complete and hence solvable.

\subsection{Back to physical space}
We have seen that $U^{(1)}_+(x)$ and $U^{(0)}_+(x)$ are completely determined by \equ{eq_symm_pair2_5} and \equ{eq_symm_pair1_6}. Moreover, from the symmetry relation \equ{eq_U_symmetry} we obtain
\begin{align}
    U^{(0)}_-(x) &= U^{(0)}_+(x^{-1}) + \operatorname{s}_p(\ell), \\
    U^{(1)}_-(x) &= U^{(1)}_+(x^{-1}) - u^*, 
\end{align}
where $u^*$ is defined in \equ{eq_unknown_1}.  
Finally, using \equ{eq_psi0}-\equ{eq_phi1}, we derive
\begin{align}
    \Psi_0(x) &= U^{(0)}_+(x^{-1}) - U^{(0)}_+(x) + \operatorname{s}_p(\ell), \label{eq_duct_sym_solution1}\\
    \Phi_1(x) &= U^{(1)}_+(x^{-1}) + U^{(1)}_+(x) - u^*. \label{eq_duct_sym_solution2}
\end{align}


We recover the scattered field in three regions:  
for $0 \leqslant n \leqslant \ell_0-1$ from  \equ{eq_transform_odd_inv} and \equ{eq_ansatz_b_sol} and the boundary condition \equ{eq_strip_bc} with \equ{eq_u_in_duct};  
for $-\ell \leqslant n < 0$, from \equ{eq_transform_even_inv} and \equ{eq_ansatz_1A3}; 
and for $\ell_0 \leqslant n \leqslant \ell_0+\ell-1$, by the symmetry of the waveguide \equ{phi_symm}-\equ{psi_symm}.  
In each case, the scattered field is obtained as follows:
\begin{alignat}{2}
    u^\text{sc}_{m,n} &= \frac{1}{2 \pi i} \oint_{\mathcal{C}} \Psi(x,n) x^{m-1} dx - 2\operatorname{s}_p(n+\ell)\delta_{m,0}, &&\quad m \geqslant 0,\ 0\leqslant n \leqslant \ell_0-1; \label{eq_sol_phys_middle_wg}
    \\   
    u^\text{sc}_{m,n} &= \frac{1}{2 \pi i} \oint_{\mathcal{C}} \Phi(x,n) x^{m-1} dx, &&\quad m>0,\  -\ell\leqslant n<0; \label{eq_sol_phys_down_wg}
    \\
    u^\text{sc}_{m,n} &= u_{m,\ell_0-1-n}, &&\quad m>0,\  \ell_0 \leqslant n\leqslant \ell_0+\ell-1.
    \label{eq_sol_phys_up_wg}
\end{alignat}

\begin{remark}
Analogously to the symmetric waveguide case, we may also obtain a solution for the system of equations \equ{eq_pair1_1}-\equ{eq_pair1_2}, \equ{eq_pair2_1}-\equ{eq_pair2_2} in the general setting via the pole removal method. A sketch of this analysis is provided in \appx{sect_4x4_pole_removal}.
\end{remark}

\begin{figure}[ht]
\centering
    \begin{subfigure}[b]{0.33\textwidth}
        \centering
        \includegraphics[height=3.5cm]{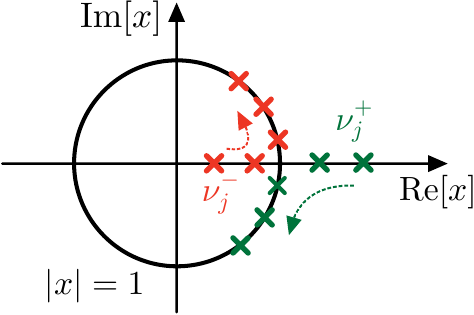}
        \caption{$\Omega^{\text{min}}_1<\Omega<\Omega^{\text{min}}_{2J-1}$.}
        \label{Omega_min}
    \end{subfigure}
    \hfill
    \begin{subfigure}[b]{0.3\textwidth}
        \centering       
        \includegraphics[height=3.5cm]{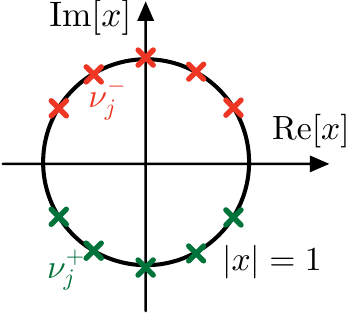}
        \caption{$\Omega^{\text{min}}_{2J-1}<\Omega<\Omega^{\text{max}}_1$.} 
        \label{Omega_2}
    \end{subfigure}
    \hfill
    \begin{subfigure}[b]{0.33\textwidth}
        \centering
        \includegraphics[height=3.5cm]{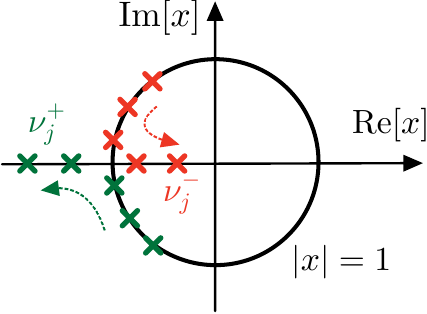}
        \caption{$\Omega^{\text{max}}_1<\Omega<\Omega^{\text{max}}_{2J-1}$.}
        \label{Omega_max}
    \end{subfigure}
\caption{\rev{Schematic representation of the position of the poles $\nu_j^\pm,$ $j\in\{1,\ldots, J\}$ of the integrands in \equ{eq_sol_phys_middle_wg}-\equ{eq_sol_phys_down_wg}, or the function in \equ{eq_system_sol2}, with respect to the unit circle for various values of $\Omega$, where $\Omega$ is strictly real. The poles lying on the unit circle correspond to the modes propagating at frequency $\Omega$, whereas the poles lying on the real axis correspond to the evanescent modes. The dashed arrows indicate the trajectory of the poles moving in the complex plane as the frequency $\Omega$ is increasing: a $(2j-1)$-th mode becomes propagating as $\Omega$ reaches the value of $\Omega^{\text{min}}_{2j-1}$ (a), and becomes evanescent again when $\Omega=\Omega^{\text{max}}_{2j-1}$ (c). For $\Omega^{\text{min}}_{2J-1}<\Omega<\Omega^{\text{max}}_1$ all the existing modes are propagating through the waveguide (b). Note that for $\Omega\in[0,\Omega^{\text{min}}_1)\cup (\Omega^{\text{max}}_{2J-1},2\sqrt{2}]$ all the poles lie on the real line and no modes are propagating in the waveguide.}}
\label{fig_Omega_poles}
\end{figure}

Note that the integrands in \equ{eq_sol_phys_middle_wg}-\equ{eq_sol_phys_down_wg} are meromorphic functions, with each pole corresponding to a waveguide mode (see \fig{fig_Omega_poles}). By the residue theorem, the amplitude of each mode in \equ{eq_sol_phys_down_wg} can be obtained from the residues at these poles:
\begin{equation}
    u^\text{sc}_{m,n}=\sum\limits_{q} M_{q}\,
    u^q_{m,n}, \qquad q=2j-1,\quad j=\{1,J\}
    \label{eq:mode_decompose}
\end{equation}
where $u^q_{m,n}$ is defined in \equ{eq_mode_form}, and the amplitude $M_q$ is given by
\begin{equation}
   M_{q} = \operatorname{Res}\left[\frac{\Phi_1(x)}{x \operatorname{s}(\ell-1)},\  x=\nu_j^-\right], \quad j \in J. \label{eq_R_res}
\end{equation}

In the indexing of \equ{eq:mode_decompose}, we account for the fact that the odd incident mode in the fully symmetric waveguide, as discussed earlier, can only be scattered into odd modes. The coefficients $M_q$ can in principle be obtained analytically from \equ{eq_symm_pair1_6} and \equ{eq_duct_sym_solution2}, although the resulting expressions are lengthy and are therefore omitted here.
\rev{\begin{remark}
Note that, as follows from the schematic representation of the pole trajectories of \equ{eq_sol_phys_middle_wg}-\equ{eq_sol_phys_down_wg} shown in \fig{fig_Omega_poles}, when the frequency $\Omega$ takes the critical values \equ{eq_omega_crit_def}-\equ{eq_omega_crit_max_def}, the poles inside and outside the unit circle coalesce into a double pole at $x=1$ for $\Omega=\Omega_j^{\text{min}}$ (\fig{Omega_min}) and at $x=-1$ for $\Omega=\Omega_j^{\text{max}}$ (\fig{Omega_max}). Although this does not affect \equ{eq_R_res}, extra care must be taken when reconstructing the physical field from the integrals \equ{eq_sol_phys_middle_wg}-\equ{eq_sol_phys_down_wg} at these critical frequencies, because it is unclear how the contour $\mathcal{C}$ should be deformed within the region of analyticity of the integrands. One possible approach to the numerical integration is to add a small imaginary part to $\Omega$, thereby shifting the poles away from the unit circle.
\end{remark}}

We have successfully implemented this residue-based method; however, in order to validate our results, we require an additional independent numerical approach. For this reason, we now turn to the Boundary Algebraic Equations method.

%% file: 5_BAE.tex
\section{Numerical solution via Boundary Algebraic Equations method}
\label{eq_bae}
The exact solution to the problem formulated in \equ{eq_Helmholtz_discrete_sc}-\equ{eq_BC_symmetric2} can be derived by the so-called Boundary Algebraic Equations (BAE) method \cite{Martinsson2009BoundaryProblems,Gillman2010FastLattices}, which can be employed for discrete scattering problems involving a finite scatterer on a square lattice \cite{Poblet-Puig2015SuppressionEquations} using the lattice Green's function. This method represents a fully discretised counterpart to the well-known boundary element method, which utilises the Green's function for continuous space, as formulated in \cite{Wickham1981TheExistence}. An exact solution for the scattering of a plane wave by a finite strip in free space was previously obtained by Sharma \cite{Sharma2015Near-tipConstraint}, and implemented in our work \cite{Medvedeva2024DiffractionMethod} for comparison. Our objective is to derive a similar result for the waveguide. To do so, we must derive an expression for the tailored lattice Green's function, which satisfies boundary conditions on the waveguide walls \equ{eq_bc_walls}.

\subsection{Tailored Green's function for the waveguide}
We introduce a tailored Green's function $G_{m,n}^{m_0,n_0}$, where $(m_0,n_0)$ is a location of a point source, such that $G_{m,n}^{m_0,n_0}$ satisfies the following:
\begin{align}
        &\Delta G_{m,n}^{m_0,n_0} + \Omega^2 G_{m,n}^{m_0,n_0} = \delta_{m,m_0}\delta_{n,n_0}, \label{eq_green_system_eq}
        \\
       & G_{m,-N_1}^{m_0,n_0}=G_{m,N_2}^{m_0,n_0} =0,
    \label{eq_green_system_bc}
\end{align}
where $\delta_{i,j}$ is the Kronecker delta, defined by \equ{eq_kronecker_def}.

We introduce the discrete Fourier transform and its inverse as:
\begin{equation}
    \widehat{G}_{n}(x)=\sum_{m=-\infty}^{\infty} G_{m,n}^{m_0,n_0} x^{-m}, \qquad G_{m,n}^{m_0,n_0} =
    \displaystyle\frac{1}{2 \pi i} \oint_{\mathcal{C}}\widehat{G}_{n}(x) x^{m-1} dx,
    \label{eq_green_fourier_transform}
\end{equation}
where the integration contour $\mathcal{C}$ lies within the region of analyticity of $\widehat{G}_{n}$, which can be taken as the unit circle for complex $\Omega$. Note that $\widehat{G}_{n}$ also depends on $m_0$ and $n_0$, but for brevity we do not display it explicitly. 

Applying \equ{eq_green_fourier_transform} to \equ{eq_green_system_eq}-\equ{eq_green_system_bc} leads to:
\begin{align}
    &\Lambda \widehat{G}_{n} + \widehat{G}_{n-1} +\widehat{G}_{n+1} = x^{-m_0}\delta_{n,n_0}, \label{eq_green_system_transformed_eq}
    \\
    &\widehat{G}_{-N_1} = \widehat{G}_{N_2}=0,
    \label{eq_green_system_transformed_bc}
\end{align}
where $\Lambda(x)$ is defined by \equ{eq_Lambda_def}. Then, we seek  a solution of \equ{eq_green_system_transformed_eq}-\equ{eq_green_system_transformed_bc} in the forms:
\begin{alignat}{2}
    &\widehat{G}_{n}^{(1)} = A_1(x)\ y^n + B_1(x)\ y^{-n}, \quad &&-N_1<n<n_0; \label{eq_G_ansatz_1}\\
    &\widehat{G}_{n}^{(2)}  = A_2(x)\ y^n + B_2(x)\ y^{-n}, \quad &&n_0<n<N_2; \label{eq_G_ansatz_2}
   \\ &\widehat{G}_{n_0} = A_0(x)\ y^{n_0} + B_0(x)\ y^{-n_0}+\frac{1}{\Lambda+2}, \quad &&n=n_0; \label{eq_G_ansatz_b}
\end{alignat}
where $y(x)$ is defined by \equ{eq_y_definition}.

Writing \equ{eq_green_system_transformed_eq} for $n=-N_1+1$ and $n=N_2-1$ and considering the boundary conditions \equ{eq_green_system_transformed_bc}:
\begin{alignat}{3}
    &\Lambda \widehat{G}_{-N_1+1}^{(1)} &&+ \widehat{G}_{-N_1+2}^{(1)} &&= 0,
    \label{eq_green_helmholts_spectral_wall_1}\\
    &\Lambda \widehat{G}_{N_2-1}^{(2)} &&+ \widehat{G}_{N_2-2}^{(2)} &&= 0.
    \label{eq_green_helmholts_spectral_wall_2}
\end{alignat}

Then, substituting \equ{eq_G_ansatz_1}-\equ{eq_G_ansatz_2} into \equ{eq_green_helmholts_spectral_wall_1}-\equ{eq_green_helmholts_spectral_wall_2}, we can eliminate $B_1$ and $B_2$ in \equ{eq_G_ansatz_1}-\equ{eq_G_ansatz_2}:
\begin{alignat}{2}
    &\widehat{G}_{n}^{(1)} = A_1^*\ \operatorname{s}(n+N_1), \qquad &&n<n_0; \label{eq_G_ansatz_1_rep}
    \\
    &\widehat{G}_{n}^{(2)}  = A_2^*\ \operatorname{s}(n-N_2), \qquad &&n>n_0; \label{eq_G_ansatz_2_rep}
\end{alignat}
where $A_1^* = A_1 y^{-N_1},$ $A_2^* = A_2 y^{N_2}$ and $\operatorname{s}(n)$ is defined by \equ{eq_s_def}.

From the equation \equ{eq_green_system_transformed_eq}, written for $n=n_0,\ n=n_0+1,\ n=n_0-1$ we obtain:
\begin{eqnarray}
\begin{cases}
    \begin{array}{lclclll}
    \Lambda \widehat{G}_{n_0}&+& \widehat{G}_{(n_0+1)}^{(2)}  &+& \widehat{G}_{(n_0-1)}^{(1)} &=& x^{-m_0}, \\
    \Lambda \widehat{G}_{(n_0-1)}^{(1)}  &+& \widehat{G}_{n_0} &+& \widehat{G}_{(n_0-2)}^{(1)} &=& 0, \\
    \Lambda \widehat{G}_{(n_0+1)}^{(2)}  &+& \widehat{G}_{(n_0+2)}^{(2)}  &+& \widehat{G}_{n_0} &=& 0. 
    \end{array}
    \label{eq_green_system_spec}
\end{cases}
\end{eqnarray}
Substituting \equ{eq_G_ansatz_1_rep}-\equ{eq_G_ansatz_2_rep} to \equ{eq_green_system_spec}, we can solve the system for $A_1^*$, $A_2^*$, $\widehat G_{n_0}$ and therefore find :
 \begin{align}
    &\widehat{G}_{n}^{(1)} = \frac{\operatorname{s}(n_0-N_2)}{\operatorname{s}(N)\ \operatorname{s}(1)}\ \operatorname{s}(n+N_1)\ x^{-m_0}, \qquad &&n<n_0; \label{eq_G1_sol}
    \\
    &\widehat{G}_{n}^{(2)}  = \frac{\operatorname{s}(n_0+N_1)}{\operatorname{s}(N)\ \operatorname{s}(1)}\ \operatorname{s}(n-N_2)\ x^{-m_0}, \qquad &&n>n_0; \label{eq_G2_sol}
    \\
    &\widehat G_{n_0} = \frac{\operatorname{s}(n_0-N_2)\ \operatorname{s}(n_0+N_1)}{\operatorname{s}(N)\ \operatorname{s}(1)}\ x^{-m_0}\qquad &&n=n_0;\label{eq_G0_sol}
 \end{align}

Note that both \equ{eq_G1_sol} and \equ{eq_G2_sol} give \equ{eq_G0_sol} when written for $n=n_0.$ Hence, the tailored Green's function for the waveguide can be recovered by the inverse transform as defined in \equ{eq_green_fourier_transform} as:
\begin{equation}
    G_{m,n}^{m_0,n_0} =
    \begin{cases}
    \displaystyle\frac{1}{2 \pi i} \oint_{\mathcal{C}}\widehat{G}_{n}^{(1)} x^{m-1} dx, \qquad -N_1<n\leqslant n_0; \\
    \displaystyle\frac{1}{2 \pi i} \oint_{\mathcal{C}}\widehat{G}_{n}^{(2)} x^{m-1} dx, \qquad n_0\leqslant n<N_2. 
    \end{cases}
    \label{eq_tailored_green_sol}
\end{equation}


\subsection{Boundary Algebraic Equations system}
The solution in the form of a system of boundary algebraic equations for the scattering by a finite Dirichlet strip in free space has been obtained in \cite{Sharma2015Near-tipConstraint}. In the case of a waveguide, the solution is analogous, apart from the fact that instead of the discrete free-space Green's function, we use the tailored lattice Green's function \equ{eq_tailored_green_sol}.

The solution for the scattered field at some point $(m_0,n_0)$ is written as:
\begin{equation}
    u^\text{sc}_{m_0,n_0} = \sum_{n=-n_1}^{n_2} G_{0,n}^{m_0,n_0} [2u^\text{sc}_{1,n} + (u_{-1,n}^{\text{in}}+u_{1,n}^{\text{in}}) + \delta_{n,n_2}(u^\text{sc}_{0,n+1}+u_{0,n+1}^{\text{in}}) + \delta_{n,-n_1}(u^\text{sc}_{0,n-1}+u_{0,n-1}^{\text{in}})].
    \label{eq_solution_BAE}
\end{equation}
We refer to \appx{appx_bae} for the detailed derivation of \equ{eq_solution_BAE}.

The coefficients $u^\text{sc}_{1,n}$ in \equ{eq_solution_BAE} are obtained from writing \equ{eq_solution_BAE} for $m_0=1, \ n_0\in\{-n_1,n_2\}$ as:
\begin{equation}
    \boldsymbol{u}=\left(\mathbf{I}_{\mathcal{N}}-\mathbf{G}_{\mathcal{N}}\right)^{-1} \left[\mathbf{G}_{\mathcal{N}} \boldsymbol{u}^{\text{in}} + (u^\text{sc}_{0,n_2+1}+u_{0,n_2+1}^{\text{in}})\mathbf{p} + (u^\text{sc}_{0,-n_1-1}+u_{0,-n_1-1}^{\text{in}})\mathbf{q}\right], \label{eq_BAE_sol}
\end{equation}
where $\mathcal{N}=n_1+n_2+1$, $\boldsymbol{u}=[u^\text{sc}_{1,-n_1},...,u^\text{sc}_{1,n_2}]^\top$, $\boldsymbol{u}^{\text{\normalfont in}}=\frac{1}{2}[(u_{-1,-n_1}^{\text{in}}+u_{1,-n_1}^{\text{in}}),...,(u_{-1,n_2}^{\text{in}}+u_{1,n_2}^{\text{in}})]^\top$, $\mathbf{I}_{\mathcal{N}}$ is the $N\times N$ identity matrix and the $\mathcal{N}\times \mathcal{N}$ matrix
$\mathbf{G}_{\mathcal{N}}$ and $\mathcal{N}\times1$ vectors $\mathbf{p}$ and $\mathbf{q}$ are defined by
\begin{equation}
    \mathbf{G}_{\mathcal{N}} =
    2\left(
    \begin{array}{llll}
     G^{1,-n_1}_{0,-n_1} & G^{1,-n_1}_{0,-n_1+1}  & \hdots & G^{1,-n_1}_{0,n_2}  \\
     G^{1,-n_1+1}_{0,-n_1} & G^{1,-n_1+1}_{0,-n_1+1} & \hdots & \vdots\\
     \vdots  & \ddots & G^{1,n_2-1}_{0,n_2-1}  & G^{1,n_2-1}_{0,n_2} \\
     G^{1,n_2}_{0,-n_1} & \hdots & G^{1,n_2}_{0,n_2-1} & G^{1,n_2}_{0,n_2} \\
    \end{array}
    \right),
    \quad
    \mathbf{p} = 
    \left(
    \begin{array}{l}
        G^{1,-n_1}_{0,n_2} \\
        G^{1,-n_1+1}_{0,n_2}\\
        \vdots \\
        G^{1,n_2}_{0,n_2}
    \end{array}
    \right),
    \quad
    \mathbf{q} =
    \left(
    \begin{array}{l}
        G^{1,-n_1}_{0,-n_1} \\
        G^{1,-n_1+1}_{0,-n_1}\\
        \vdots \\
        G^{1,n_2}_{0,-n_1}
    \end{array}
    \right).
\end{equation}
The two unknown constants $u^\text{sc}_{0,-n_1-1}$ and $u^\text{sc}_{0,n_2+1}$ in \equ{eq_solution_BAE} and \equ{eq_BAE_sol} are found after some simple linear algebraic transformations from \equ{eq_solution_BAE}, written for $m_0=0,\ n_0=-n_1-1$ and $m_0=0,\ n_0=n_2+1$, as:
\begin{equation}
    (\mathbf{I}_2 - \mathbf{G}_2- 2\mathbf{H}_2)\boldsymbol{u}^{*} = (\mathbf{G}_2+2\mathbf{H}_2)\boldsymbol{u}^{*\text{\normalfont in}}  + 2\left( 
    \begin{array}{cc}
         \mathbf{g}(\mathbf{I}_{\mathcal{N}} + \left(\mathbf{I}_{\mathcal{N}}-\mathbf{G}_{\mathcal{N}}\right)^{-1}\mathbf{G}_{\mathcal{N}}) \boldsymbol{u}^{\text{\normalfont in}}  \\
         \mathbf{h}(\mathbf{I}_{\mathcal{N}} + \left(\mathbf{I}_{\mathcal{N}}-\mathbf{G}_{\mathcal{N}}\right)^{-1}\mathbf{G}_{\mathcal{N}}) \boldsymbol{u}^{\text{\normalfont in}}
    \end{array}
    \right), 
\end{equation}
where $\mathbf{I}_2$ is the $2\times2$ identity matrix, and
\begin{equation*}
    \boldsymbol{u}^{*}=\left(
    \begin{array}{l}
        u^\text{sc}_{0,-n_1-1}  \\
        u^\text{sc}_{0,n_2+1}
    \end{array}\right),
    \qquad
    \boldsymbol{u}^{*\text{in}}=\left(
    \begin{array}{l}
        u_{0,-n_1-1}^{\text{in}}  \\
        u_{0,n_2+1}^{\text{in}}
    \end{array}\right),
\end{equation*}
\begin{equation*}
    \mathbf{H}_2 = \left( 
    \begin{array}{ccc}
         \mathbf{g}\left(\mathbf{I}_{\mathcal{N}}-\mathbf{G}_{\mathcal{N}}\right)^{-1}\mathbf{q} & & \mathbf{g}\left(\mathbf{I}_{\mathcal{N}}-\mathbf{G}_{\mathcal{N}}\right)^{-1}\mathbf{p}  \\
         \mathbf{h}\left(\mathbf{I}_{\mathcal{N}}-\mathbf{G}_{\mathcal{N}}\right)^{-1}\mathbf{q} & & \mathbf{h}\left(\mathbf{I}_{\mathcal{N}}-\mathbf{G}_{\mathcal{N}}\right)^{-1}\mathbf{p}
    \end{array}
    \right),
    \quad 
    \mathbf{G}_2 = \left(
    \begin{array}{ll}
        G^{0,-n_1-1}_{0,-n_1} & G^{0,-n_1-1}_{0,n_2}  \\
        G^{0,n_2+1}_{0,-n_1} & G^{0,n_2+1}_{0,n_2}
    \end{array}
    \right),
\end{equation*}
\begin{align*}
    \mathbf{g} = [G^{0,-n_1-1}_{0,-n_1}, ...,G^{0,-n_1-1}_{0,n_2}], \qquad \mathbf{h} = [G^{0,n_2+1}_{0,-n_1}, ...,G^{0,n_2+1}_{0,n_2}].
\end{align*}

\begin{figure}[ht]
\centering
    \begin{subfigure}[b]{0.49\textwidth}
        \centering
        \caption{Re$[u^\text{in}_{m,n}]$ for $\Omega=0.5$.}
        \includegraphics[width=\textwidth]{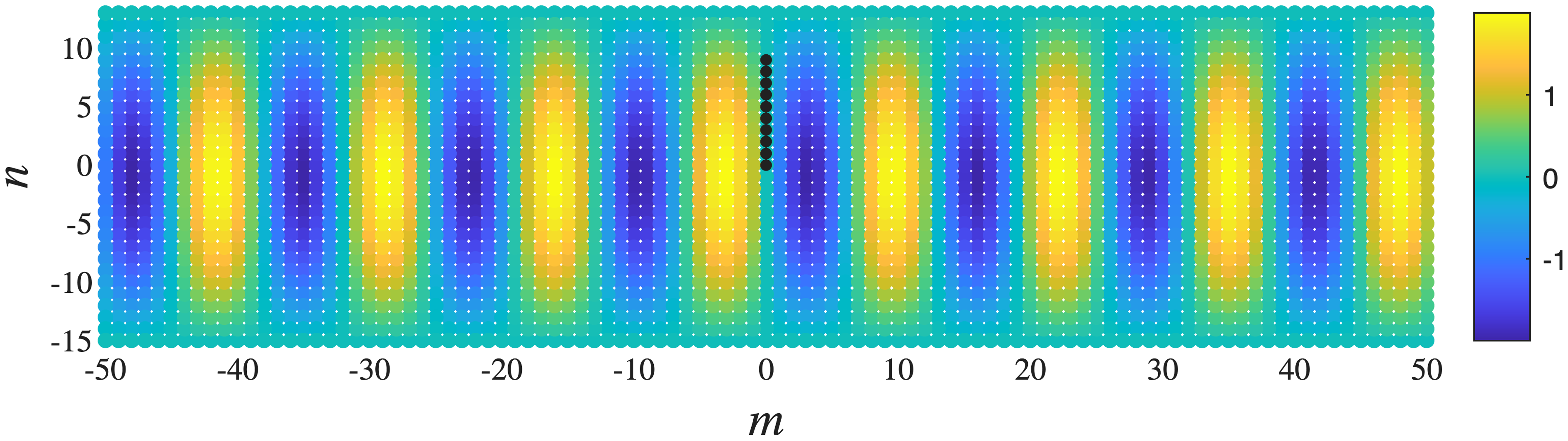}
        \label{bae_0_5_in}
    \end{subfigure}
    \begin{subfigure}[b]{0.49\textwidth}
        \centering
        \caption{Re$[u^\text{in}_{m,n}]$ for $\Omega=1.5$.}        
        \includegraphics[width=\textwidth]{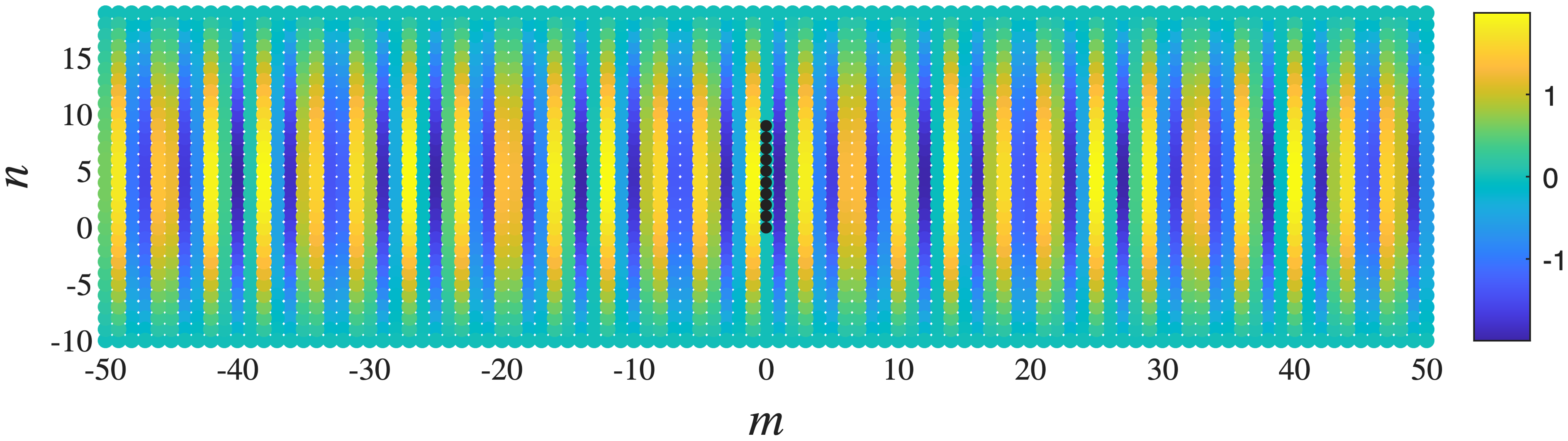}
        \label{bae_1_5_in}
    \end{subfigure}
    \begin{subfigure}[b]{0.49\textwidth}
        \centering
        \caption{Re$[u^\text{sc}_{m,n}]$ for $\Omega=0.5$.}
        \includegraphics[width=\textwidth]{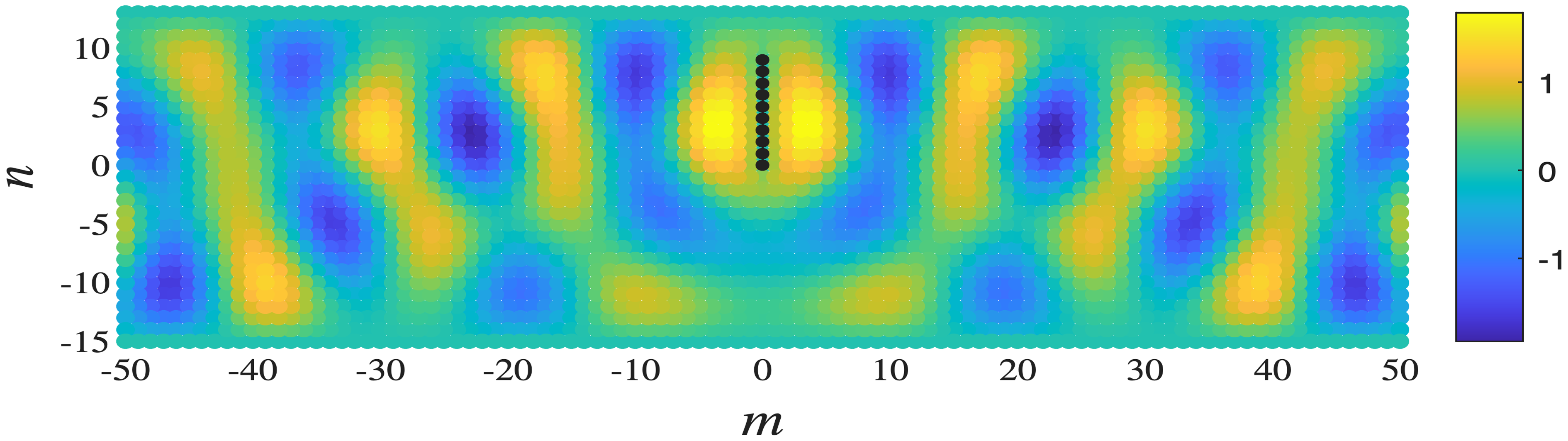}
        \label{bae_0_5_sc}
    \end{subfigure}
    \begin{subfigure}[b]{0.49\textwidth}
        \centering
        \caption{Re$[u^\text{sc}_{m,n}]$ for $\Omega=1.5$.}         \includegraphics[width=\textwidth]{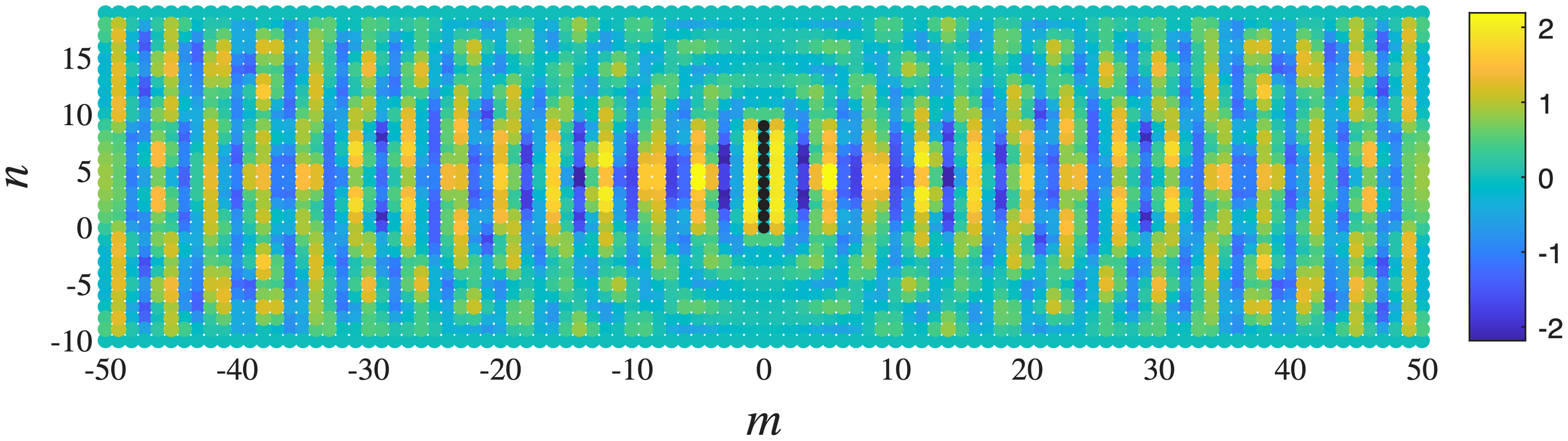}
        \label{bae_1_5_sc}
    \end{subfigure}
    \begin{subfigure}[b]{0.49\textwidth}
        \centering
        \caption{Re$[u^\text{tot}_{m,n}]$ for $\Omega=0.5$.}
        \includegraphics[width=\textwidth]{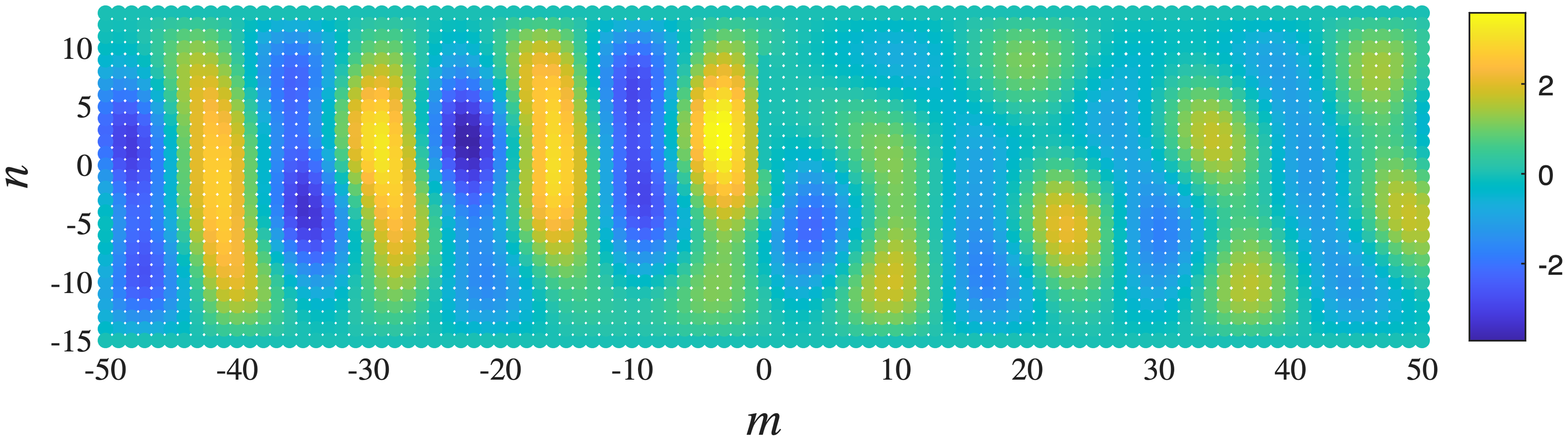}
        \label{bae_0_5_tot}
    \end{subfigure}
    \begin{subfigure}[b]{0.49\textwidth}
        \centering
        \caption{Re$[u^\text{tot}_{m,n}]$ for $\Omega=1.5$.}         \includegraphics[width=\textwidth]{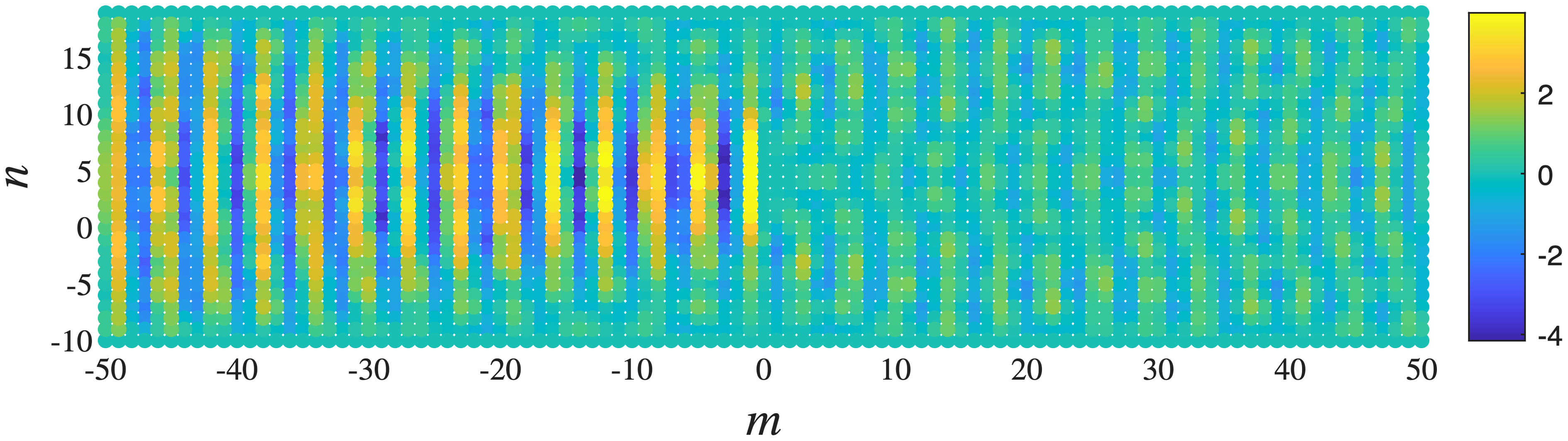}
      \label{bae_1_5_tot}
    \end{subfigure}
\caption{Real part of the incident, scattered and total fields for the discrete waveguide with a transversal Dirichlet screen, where the number of the incident mode is $p=1$, defined by \equ{eq_u_in_duct}, for two configurations: $\Omega=0.5$, $N_1=15$, $N_2=13$, $n_1=0, n_2=9$  (a), (c), (e) and  $\Omega=\rev{1.5}$, $N_1=10$, $N_2=19$, $n_1=0, n_2=9$ (b), (d), (f). The location of the screen is marked by black dots for the incident and scattered fields (a)-(d). In figures (e) and (f), one can observe that the Dirichlet boundary condition on the screen is satisfied.}
\label{fig:bae_strip}
\end{figure}

The resulting wavefield, obtained numerically from the solution \equ{eq_BAE_sol} for incident mode $p=1$, is shown in \fig{fig:bae_strip}. Note that we have also recovered the wavefield for the symmetric waveguide, as in \fig{bae_1_5_sc} and \fig{bae_1_5_tot}, by numerically evaluating the integral solutions from the pole removal method \equ{eq_sol_phys_middle_wg}--\equ{eq_sol_phys_down_wg} together with \equ{eq_sol_phys_up_wg}. The absolute error between the two methods is about $10^{-11}$, which corresponds to the numerical accuracy of the integral evaluation in \texttt{MATLAB}.

\begin{remark}
 Note that, in both solution methods considered in \sect{sect_pole_removal} and \sect{eq_bae}, the problem is reduced to solving a finite system of linear equations. However, the sizes of these systems differ. In the pole removal method, the system size $J$ is determined by the length of the gap and the half-length of the screen,
$
J = \left\lfloor \tfrac{\ell_0}{2} \right\rfloor + (\ell - 1),
$
whereas in the boundary algebraic equations method the size of the system is equal only to the length of the screen, namely $\ell_0=n_1+n_2+1$.
\end{remark}

%% file: 5_0_RT.tex
\FloatBarrier
\section{Reflection and transmission coefficients}
\label{sect:future}
To find the reflection and transmission coefficients, we represent the total field on the left and right sides of the strip as:
\begin{equation}
    u^\text{tot}_{m,n}=u^\text{sc}_{m,n}+u^\text{in}_{m,n}=
    \begin{cases}
        u^\text{in}_{m,n}+u^\text{ref}_{m,n}, &m<0;
        \\
        u^\text{tr}_{m,n}, &m>0;
    \end{cases}
    \label{eq_RT_def_tot}
\end{equation}
where $u^\text{ref}_{m,n}$ is the reflected field and $u^\text{tr}_{m,n}$ is the transmitted field. From  \equ{eq:mode_decompose}, \equ{eq_RT_def_tot} and the symmetry property of the scattered field \equ{eq_field_symmetry} if follows that
\begin{align}
    &u^\text{ref}_{m,n} = u^\text{sc}_{m,n}=\sum\limits_{q} M_{q}\,
    u^q_{-m,n}, && m<0,
    \label{eq_def_ref}
    \\
    &u^\text{tr}_{m,n}  = u^\text{sc}_{m,n} + u^\text{in}_{m,n}=\sum\limits_{q} M_{q}\,
    u^q_{m,n}+u^p_{m,n}, && m>0,
    \label{eq_def_utr}
\end{align}
where $q$ and $u^q_{m,n}$ are defined in \equ{eq_mode_form}.

From \equ{eq_def_ref}--\equ{eq_def_utr} it follows that the amplitudes of the scattered modes $M_q$, defined by \equ{eq_R_res}, correspond to the reflection coefficients:
\begin{equation}
    R_q = M_q. 
\end{equation}
The transmission coefficient for the incident mode is
\begin{equation}
    T_{p} = 1 + M_p, 
    \label{eq_RT_def}
\end{equation}
where $p$ denotes the index of the incident mode, while the transmission coefficients for the other supported modes, with $q \neq p$, are 
\begin{equation}
    T_q = M_q, \quad q \neq p.
    \label{eq_RT_def2}
\end{equation}

Numerically, $T_q$ and $R_q$ can also be obtained from the wavefield, computed using \equ{eq_solution_BAE} or \equ{eq_sol_phys_middle_wg}-\equ{eq_sol_phys_up_wg} by exploiting the orthogonality of the waveguide modes. 

To illustrate what we have in mind by orthogonality, we define the discrete inner product by
\begin{equation}
\langle f,g\rangle = \sum_{n=-N_1}^{N_2} f(n)\,\overline{g(n)}, 
\label{eq_inner_product_def}
\end{equation}
with the overbar denoting complex conjugation.
For all mode numbers $1 \leqslant j,k \leqslant N-1$, \equ{eq_mode_form} gives
\[
\langle \operatorname{s}_j,\operatorname{s}_k\rangle = 2\rev{N}\,\delta_{jk}, 
\qquad 
\|\operatorname{s}_j\|^2 = \langle \operatorname{s}_j,\operatorname{s}_j\rangle = 2\rev{N}.
\]

Then, for some $m_r \gg 1$, which is a distance between some waveguide cross-section and the strip, the transmission and reflection coefficients for mode number $q \in \{1,\dots,J\}$ are given by
\begin{equation}
T_q = \frac{\langle u^\text{tot}_{m_r,n},\,\operatorname{s}_q\rangle}{\|\operatorname{s}_q\|^2}\,x_q^{\,m_r},
\qquad
R_q = \frac{\langle u^\text{sc}_{-m_r,n},\,\operatorname{s}_q\rangle}{\|\operatorname{s}_q\|^2}\,x_q^{\,m_r}, 
\label{eq_RT_numerical}
\end{equation}
where the incident mode $p$ is chosen to be of unit amplitude, meaning that
\[
\frac{\langle u^\text{in}_{0,n},\,\operatorname{s}_p\rangle}{\|\operatorname{s}_p\|^2} = 1.
\]

\begin{figure}[htbp]
\centering
    \includegraphics[width=0.4\textwidth]{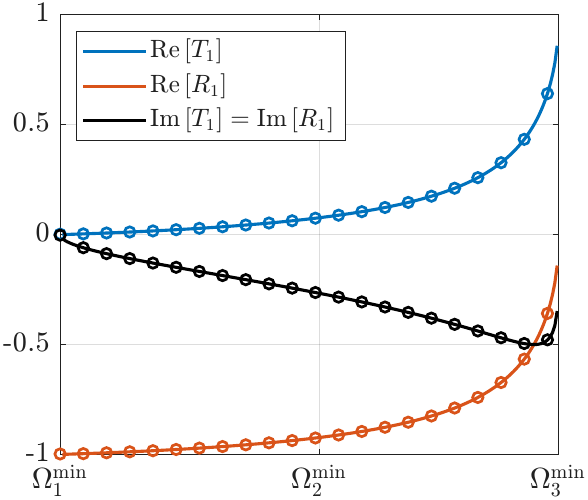}
    \caption{Real parts $\text{Re}[T_1]$ and $\text{Re}[R_1]$ as functions of the lattice frequency $\Omega$ for \rev{$\ell=3$} and \rev{$\ell_0=3$}, with incident mode $p=1$. The black line shows the imaginary part $\text{Im}[T_1]=\text{Im}[R_1]$. The solid line shows the analytic results obtained via the residue theorem from \equ{eq_R_res} and \equ{eq_RT_def}, while the circles denote the numerical coefficients computed by the inner products in \equ{eq_RT_numerical} from the wavefield obtained by the boundary algebraic equations system in \equ{eq_solution_BAE}.}
    \label{fig_real_RT}
\end{figure}

\fig{fig_real_RT} shows how the reflection and transmission coefficients of the incident mode number $p=1$, $\rev{T_1}$ and $\rev{R_1}$, depend on the lattice frequency $\Omega$. In particular, the reflection coefficient tends to $-1$ as $\Omega$ approaches the cut-off frequency of the incident mode $\rev{\Omega_1^{\text{min}}}$, defined in \equ{eq_omega_crit_def}. Physically, this corresponds to full reflection and zero transmission. This behaviour is consistent with the theoretical result obtained for the continuous waveguide with a transversal Dirichlet screen, as discussed in \cite{Shanin2017DiffractionWaveguideb}. \rev{It can be shown in a similar manner that $\text{Re}[R_1]\to-1,$ $\text{Re}[T_1]\to0,$ as $\Omega\to\Omega^{\text{max}}_1$.}

Note that throughout this work we have studied the incident mode with index $p=1$ as the most natural physical choice. However, the theoretical results are derived for the general case, where $p$ can take any value in $\{1,\rev{\ldots,N-1}\}$.

The total \textit{energy flux} through a transverse cross-section of the waveguide must remain constant to satisfy energy balance. The energy flux from the cross-section at $x=m$ to $x=m+1$ is given in \cite{Brillouin1946WaveLattices} as the real part of the negative inner product \equ{eq_inner_product_def} of the force applied to the cross-section $x=m$ by the cross-section $x=m+1$, and the velocity of the cross-section $x=m$:
\begin{equation}
    \mathcal{E}_f=-\frac{1}{2}\text{Re}\left[\left\langle (u^\text{tot}_{m+1,n}-u^\text{tot}_{m,n}),\ \dot u^\text{tot}_{m,n}\right\rangle\right],
    \label{eq_energy_flux_ip_def}
\end{equation} 
where $\dot u^\text{tot}_{m,n}$ in the orthogonal product denotes the time derivative of the displacement field, which is equivalent to $i \Omega u^\text{tot}_{m,n}$ in the time-harmonic case, considered in this work.

Then, writing \equ{eq_energy_flux_ip_def} for some $m<0$ and $m>0$ (on the left- and right-hand sides of the screen, respectively), using \equ{eq_RT_def_tot}--\equ{eq_def_utr} together with the inner product definition \equ{eq_inner_product_def} and the orthogonality properties of the waveguide modes, and equating the resulting expressions, after some straightforward algebraic manipulations, we obtain the energy flux balance equation:
\begin{equation}
    |v_p| - \sum_{q} |v_q|\,|R_q|^2 = \sum_{q} |v_q|\,|T_q|^2.
    \label{eq_energy_balance0}
\end{equation}
Here, $v_q$ denotes the \textit{group velocity} \cite{Brillouin1946WaveLattices, Sharma2016WaveStrips} of the mode with index $q$, which quantifies the speed at which the envelope of a wave packet, and hence the energy, propagates through the lattice waveguide. It is defined as
\begin{equation}
    v_q = \frac{d\Omega}{dK_q} = \frac{\sin K_q}{\Omega}.
    \label{eq_group_velosity_def}
\end{equation}
Note that the $q$-th waveguide mode in \equ{eq_mode_form}--\equ{eq_theta} is defined such that for positive $K_q$ the mode propagates in the positive direction of the $m$-axis. Therefore, using \equ{eq_group_velosity_def}, the group velocity of the mode with index $q$ propagating to the right is taken as $|v_q|$, while for the same mode propagating to the left it is $-|v_q|$.

\begin{figure}[ht]
\centering
    \includegraphics[width=\textwidth]{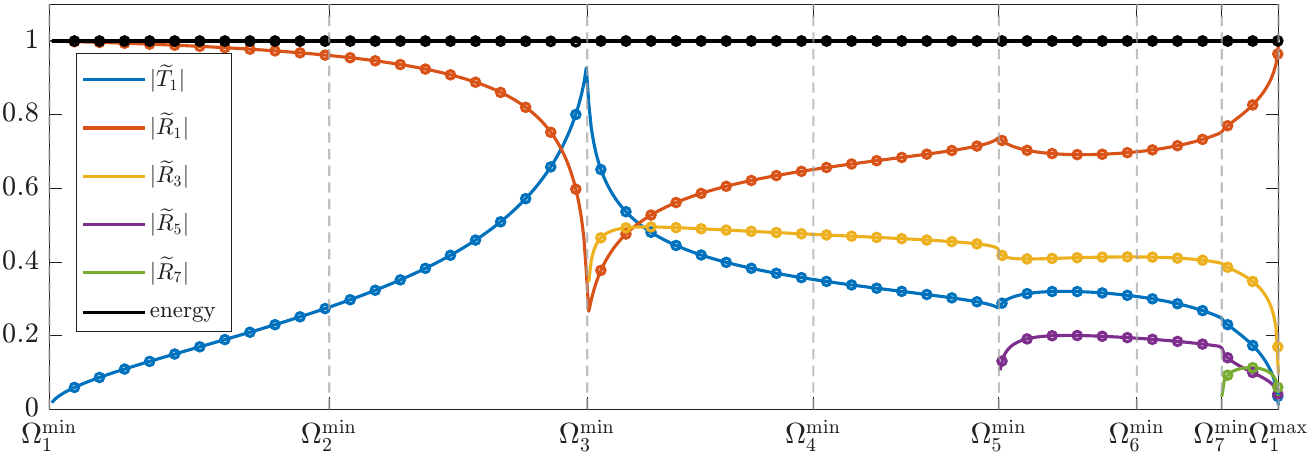}
    \caption{Frequency dependence of the magnitudes of \equ{eq_RT_def_weight} for \rev{$\ell=3$}, \rev{$\ell_0=3$}, with incident mode $p=1$. The solid line shows the analytic results obtained via the residue theorem from \equ{eq_R_res} and \equ{eq_RT_def}, while the circles denote the numerical coefficients computed by the inner products in \equ{eq_RT_numerical} from the wavefield obtained by the boundary algebraic equations system in \equ{eq_solution_BAE}. The black line shows the total energy flux.}
\label{fig_K_phaseplot}
\end{figure}

Normalising \equ{eq_energy_balance0} by $|v_p|$, we obtain the energy balance equation
\begin{equation}
    \sum_{q}\left(\frac{|v_q|}{|v_p|}\,|T_q|^2 + \frac{|v_q|}{|v_p|}\,|R_q|^2\right) = 1 \quad \text{or} \quad  \sum_{q}\left(|\widetilde{T}_q|^2 + |\widetilde{R}_q|^2\right) = 1,
    \label{eq_energy_balance}
\end{equation}
where we defined the weighted reflection and transmission coefficients by rescaling \equ{eq_RT_def} as
\begin{equation}
    |\widetilde{T}_q| = \sqrt{\frac{|v_q|}{|v_p|}}|T_q|, \qquad
    |\widetilde{R}_q| = \sqrt{\frac{|v_q|}{|v_p|}}|R_q|.
    \label{eq_RT_def_weight}
\end{equation}
The dependence of the magnitudes of \equ{eq_RT_def_weight} on the lattice frequency $\Omega$ for \rev{the four existing odd modes for the waveguide with $\ell=3$, $\ell_0=3$} is shown in \fig{fig_K_phaseplot}. Note that since the group velocity vanishes near the band edges ($K_q \to 0$ or $K_q \to \pi$) for all $q \neq 1$ 
it follows that $\widetilde R_q\to0$ and $\widetilde T_q\to0$ as $\Omega\to\rev{\Omega_q^{\text{min}}}$. \rev{Similarly, the transmission coefficients for all existing modes tend to $0$ as $\Omega\to\Omega^{\text{max}}_1$, whereas $|\widetilde{R}_1|\to1$.}


The equation \equ{eq_energy_balance} provides an additional means to verify our results. It is satisfied by the exact analytical solution in \equ{eq_R_res} up to an accuracy of $10^{-13}$, which is a remarkable result. 
As seen in \fig{fig_K_phaseplot}, the same equation for the coefficients, obtained via orthogonal products from the wavefield computed using the boundary algebraic equations method \equ{eq_solution_BAE}, is satisfied with a larger error near the cut-off frequencies than when applying the residue theorem (up to $10^{-3}$).
 This arises because we chose $m_r = 20$, which is too close to the scatterer in the considered geometry, so that some energy still leaks into the evanescent modes at the chosen cross-section. Such difficulties are not uncommon, as numerical methods often struggle near cut-off frequencies in the continuous case \cite{Lynott2019AcousticFunction}. The error can be reduced by increasing $m_r$, although here it was intentionally kept small to highlight the differences.
 
\begin{remark}
    A similar analysis can be carried out for the case of a discrete analogue of the Neumann boundary conditions for a screen in the waveguide. In this case, one expects to observe anomalous transmission \cite{Shanin2017DiffractionWaveguideb}, where $|T_1| \to 1$ and $|R_1| \to 0$ as $\Omega \to \rev{\Omega_1^{\text{min}}}$, the cut-off frequency defined in \equ{eq_omega_crit_def}.
\end{remark}

%% file: 5_conclusions.tex
\FloatBarrier
\section{Conclusions}
\label{sect_conclusion}
In this paper, we have studied wave scattering by a finite transversal strip in a discrete square-lattice waveguide, viewed as a discrete analogue of the classical continuous waveguide problem with a screen. We first formulated the boundary-value problem and then derived its associated matrix Wiener-Hopf equation.

In the general setting, this leads to a $4\times 4$ matrix kernel, while in the symmetric case it reduces to a $2\times 2$ kernel, similarly to the continuous formulation. For the Wiener--Hopf method, we only analysed the symmetric case in detail. We discussed possible approaches to the factorisation of the kernel; direct factorisation proved impractical, as the Khrapkov--Daniele form in the special case was shown to be more complicated to factorise than in the continuous case. However, the rational structure of the matrix allowed us to apply the pole removal technique. This provided an exact analytical solution, in contrast to the continuous case, where approximations are required due to the infinite number of poles in the matrix. Note that the same technique is applicable to the system with a $4\times 4$ matrix kernel (see \appx{sect_4x4_pole_removal}).

The exact solution for the wavefield, obtained by the pole removal technique, was verified numerically by applying the Boundary Algebraic Equations method with a lattice Green’s function tailored to the waveguide. The two methods provide consistent solutions with good agreement. 

From the analytical solution of the Wiener--Hopf system we derived expressions for the reflection and transmission coefficients. These were also computed numerically, via orthogonality relations applied to the wavefield obtained from the system of boundary algebraic equations, and shown to satisfy the energy balance condition. The analytical results agree with high accuracy, while the numerical results converge when the sampling is taken sufficiently far from the scatterer. Comparison with the continuous waveguide case confirms consistency of behaviour, especially near cut-off frequencies. In particular, the full reflection and zero transmission were recovered for the frequency approaching the cut-off value for the incident mode.

Overall, this work establishes a framework for treating discrete waveguide scattering problems with transversal defects, using numerical and analytical approaches and clarifying the relationship between the discrete and continuous formulations of the Wiener--Hopf method. Future work may include constructing a matrix factorisation of the matrix Wiener--Hopf kernel, as discussed in \sect{sect_matrix_fact}, where we observed that the matrix of Khrapkov--Daniele type arising in the symmetric case is more complex than its continuous analogue, yet potentially factorisable by methods such as those in \cite{Antipov2002FactorizationTheory}. Further directions include a detailed analysis of the applicability of the method from \cite{Adukov2022AnPackage} for the factorisation of matrix polynomials. While the direct use of the algorithmic software developed in \cite{Adukov2022AnPackage} appears challenging, it may be possible to apply explicitly the algorithms for the direct factorisation of meromorphic matrix functions described in \cite{Adukov1991OnFunctions}. 
It would also be worthwhile to compare the accuracy and performance of the pole-removal method for solving the diffraction problem, as discussed in \sect{sect_pole_removal}, with the more general approach of factorising the associated rational matrix. Additional developments might also consider generalising the techniques discussed here to other lattice geometries, such as those in \cite{Sharma2016EdgeSection}, motivated by physical applications, as well as exploring more complicated boundary conditions and obstacle geometries.

%% file: append.tex
\input{append_symmetry}
\input{append_cheb}

\input{append_limits}

\input{append_4x4}
\input{append_bae}

%% file: append_symmetry.tex
\section{Symmetry of the problem}
\label{sect:app_symmetry}
To analyse the symmetry properties of the problem, formulated in \sect{sect:waveguide}, we write $u^\text{in}_{m,n}$ as a sum of symmetric $u^{\text{in};\,s}_{m,n}$ and antisymmetric $u^{\text{in};\,a}_{m,n}$ components:
\begin{equation*}
    u^\text{in}_{m,n} = u^{\text{in};\ s}_{m,n} + u^{\text{in};\ a}_{m,n},
\end{equation*}
where
\begin{align}
    u^{\text{in};\ s}_{m,n} &= \frac{1}{2}(u^\text{in}_{m,n} + u^\text{in}_{-m,n}), \label{eq_u_in_sym}
    \\
    u^{\text{in};\ a}_{m,n} &= \frac{1}{2}(u^\text{in}_{m,n}-u^\text{in}_{-m,n}). \label{eq_u_in_asym}
\end{align}
Then we can split the original problem for $u^\text{sc}_{m,n}$ into two subproblems, for symmetric and anti-symmetric solutions:
\begin{alignat*}{3}
    u^{\text{sc};\ s}_{m,n} &= \frac{1}{2}(u^\text{sc}_{m,n}+u^\text{sc}_{-m,n}) &&= &&u^{\text{sc};\ s}_{-m,n},
    \\
    u^{\text{sc};\ a}_{m,n} &= \frac{1}{2}(u^\text{sc}_{m,n}-u^\text{sc}_{-m,n}) &&= -&&u^{\text{sc};\ a}_{-m,n}.
\end{alignat*}

From the boundary condition on the strip \equ{eq_strip_bc}, it follows that:
\begin{alignat*}{2}
    u^{\text{sc};s}_{0,n} &= \frac{1}{2}(u^\text{sc}_{0,n}+u^\text{sc}_{0,n}) &&= u^\text{sc}_{0,n},
    \\
    u^{\text{sc};a}_{0,n} &= \frac{1}{2}(u^\text{sc}_{0,n}-u^\text{sc}_{0,n}) &&= 0.
\end{alignat*}

\rev{
The anti-symmetric component therefore satisfies the homogeneous discrete Helmholtz equation with homogeneous boundary conditions. Moreover, it satisfies the same radiation condition as the scattered field, specified through the limiting absorption principle. 
It can be shown, similarly to \cite{Shanin2022DiffractionLattice,Kapanadze2018ExteriorLattice}, that the corresponding solution is unique, and therefore the anti-symmetric component must vanish identically:
\[
u^{\text{sc};a}_{m,n}\equiv 0.
\]
Hence, the scattered field is symmetric in $m$:
\begin{equation}
u^\text{sc}_{m,n}=u^\text{sc}_{-m,n}.
\label{eq_field_symmetry_app}
\end{equation}
}



%% file: append_cheb.tex
\section{On the roots of Chebyshev polynomials and their 
combinations}
\label{sect:cheb_roots}
By definition, the Chebyshev polynomial of the second kind and of order $J$ satisfies
\begin{equation}
    \U_J(\cos{\theta}) \sin{\theta} = \sin{(J+1)\theta}. \label{eq_chebyshev_pol_def}
\end{equation}
It follows that there are exactly $J$ roots of $\U_J(z)$ in the interval $z \in [-1,1]$, where $z = \cos{\theta}$:
\begin{equation}
    z_j = \cos{\left(\frac{\pi j}{J+1}\right)}, \qquad j \in \{1,\dots,J\}. \label{z_root}
\end{equation}

Similarly, the roots of the term $1-\U_{J}(z)$ in \equ{eq_K1_1}, with $z = \cos{\theta}$, are obtained by solving the trigonometric equation
\begin{eqnarray}
    \frac{\sin{\theta}-\sin{(J+1)\theta}}{\sin{\theta}} = 0, \qquad \theta \in (0,\pi). 
    \label{eq_cheb_roots1}
\end{eqnarray}
Equation \equ{eq_cheb_roots1} has exactly $J$ roots in the interval $(0,\pi)$, and the solutions in the interval $z \in [-1,1]$ are given by:
\begin{align}
    z_{1,j} &= 
    \cos{\left(\frac{2\pi j}{J}\right)}, &&j\in\mathbb{Z}\cap\left[1,\ J/2\right); \label{eq_chebyshev_pol_def1} 
    \\
    z_{2,j} &=
    \cos{\left(\frac{\pi (2j+1)}{J+2}\right)}, && j\in\mathbb{Z}\cap\left[0,\ (J+1)/2\right). \label{eq_chebyshev_pol_def2}
\end{align}

The roots in the complex plane $x$ are obtained by expressing $x$ from \equ{eq_z_x} as
\begin{equation}
    x_{i,\pm} = -\frac{\gamma_i}{2} \pm \frac{\sqrt{\gamma_i^2-4}}{2}, 
    \qquad \gamma_i = 2\rev{z_i} + \Omega^2 - 4, 
    \label{eq_back_to_x}
\end{equation}
where $i=\{1,\ldots,\text{(number of roots)}\}$. The branch of the square root is chosen so that $x_i^{+}$ lies outside the unit circle and $x_i^{-}$ lies inside the unit circle. Note that $x_i^{+}$ and $x_i^{-}$ are reciprocal.

Thus, we find the zeros $\mu^{(0)}_{j,\pm}$ and poles  $ \lambda^{(0)}_{j,\pm}$ of $\mathcal{K}^{(0)}(x)$ in \equ{eq_K1_1} from \equ{z_root} and \equ{eq_back_to_x} as:
\begin{align}
    \mu^{(0)}_{j,\pm} = -\frac{\gamma_j^{(0)}}{2} \pm \frac{\sqrt{\left(\gamma_j^{(0)}\right)^2-4}}{2}, 
    \qquad \gamma_j^{(0)} = 2\rev{\cos\left(\frac{\pi j}{\ell-1}\right)} + \Omega^2 - 4, \label{eq_mu0_def}
\end{align}
where $j\in\{1,\dots,\ell-2\}$, and
\begin{align}
    \lambda^{(0)}_{j,\pm} = -\frac{\zeta_j^{(0)}}{2} \pm \frac{\sqrt{\left(\zeta_j^{(0)}\right)^2-4}}{2}, 
    \qquad \zeta_j^{(0)} = 2\rev{\cos\left(\frac{\pi j}{\ell}\right)} + \Omega^2 - 4, \label{eq_lambda0_def}
\end{align}
where $j\in\{1,\dots,\ell-1\}$.

To find the zeros $\mu^{(1)}_{j,\pm}$ and poles  $ \lambda^{(1)}_{j,\pm}$ of $\mathcal{K}^{(1)}(x)$ in \equ{eq_K1_1} we note that in this case the poles found from \equ{eq_chebyshev_pol_def1} are cancelled by some of the zeros from \equ{z_root}. Thus, from \equ{z_root}, \equ{eq_chebyshev_pol_def2} and \equ{eq_back_to_x} we find:
\begin{align}
    \mu^{(1)}_{j,\pm} &= -\frac{\gamma_j^{(1)}}{2} \pm \frac{\sqrt{\left(\gamma_j^{(1)}\right)^2-4}}{2}, 
    \qquad \gamma_j^{(1)} = 2\rev{\cos\left(\frac{\pi j}{\ell_0-1}\right)} + \Omega^2 - 4, \label{eq_mu1_def}
\end{align}
where $$j\in\mathbb{Z}\cap\left(\frac{\ell_0-2}{2},\ell_0-2\right],$$
and
\begin{align}
    \lambda^{(1)}_{j,\pm} &= -\frac{\zeta_j^{(1)}}{2} \pm \frac{\sqrt{\left(\zeta_j^{(1)}\right)^2-4}}{2}, 
    \qquad \zeta_j^{(1)} = 2\rev{\cos\left(\frac{\pi (2j+1)}{\ell_0+1}\right)} + \Omega^2 - 4, \label{eq_lambda1_def}
\end{align}
where $$j\in\mathbb{Z}\cap\left[0,\ \frac{\ell_0}{2}\right).$$

To find the poles of \equ{eq_system_sol2}, we need to find the roots of the denominator, i.e.\ solve the trigonometric equation:
\begin{equation}
    \frac{(\sin{\theta}-\sin{\ell_0\theta})\sin{\ell\theta}+\sin{((\ell_0-1)\theta)}\sin{((\ell-1)\theta})}{\sin{\theta}}=0,
\end{equation}
which simplifies to
\begin{equation}
    \sin{\ell\theta}-\sin{((\ell+\ell_0-1)\theta)}=0. \label{eq_trig_poles}
\end{equation}
The solutions of \equ{eq_trig_poles} in the interval $(0,\pi),$ and the solutions in the interval $z \in [-1,1]$ are given by:
\begin{align}
    &z_{3,j} = \cos \left(\frac{2\pi j}{\ell_0-1}\right), & &0< j<\frac{\ell_0-1}{2}, \label{eq_z3j_def}\\
    &z_{4,j} = \cos \left(\frac{\pi (2j+1)}{2\ell+\ell_0-1}\right), & &0\leq j<\frac{\ell_0}{2}+(\ell-1). \label{eq_z4j_def}
\end{align}
Note that \equ{eq_z3j_def} coincide with the zeros of \equ{eq_system_sol2}, determined by \equ{eq_chebyshev_pol_def1}. Thus, $\nu^{\pm}_j$ in \equ{eq_V} are defined from \equ{eq_z4j_def} and \equ{eq_back_to_x} as:
\begin{align}
    \nu^{\pm}_j &= -\frac{\xi_j}{2} \pm \frac{\sqrt{\xi_j^2-4}}{2}, 
    \qquad \xi_j = 2\rev{\cos \left(\frac{\pi (2j+1)}{2\ell+\ell_0-1}\right)} + \Omega^2 - 4, \label{eq_nupm_def}
\end{align}
where 
$$0\leq j<\frac{\ell_0}{2}+(\ell-1).$$

%% file: append_limits.tex
\section{Limits of some important functions from \sect{sect_add_split}}
\label{appx_limits}
\FloatBarrier
To apply Liouville's theorem to find a solution to the system \equ{eq_symm_pair2_22}-\equ{eq_symm_pair1_22}, we need to investigate the growth of the `$-$' and `$+$' parts of the equation at $\infty$ and $0$, for $s\in\{0,1\}$. From the definitions \equ{eq_U0_def}-\equ{eq_U1_def}, \equ{eq_def_Ks}, \equ{eq_split_term_neg}-\equ{eq_split_term_pm} and \equ{eq_ref_limits1}-\equ{eq_ref_limits2} it is reasonably straightforward to obtain: 
{%
\allowdisplaybreaks
\begin{align*}
     &\lim_{x\to0}\mathcal{K}^{(s)} = 0;& &\lim_{x\to\infty}\mathcal{K}^{(s)} = 0 ; 
     \\
     &\lim_{x\to0}\Pi_- = 0; &    &\lim_{x\to\infty}\Pi_+ = 0 ; 
     \\
     &\lim_{x\to0}U^{(0)}_- = 0; & &\lim_{x\to\infty} U^{(0)}_+ = -\operatorname{s}_p(\ell) ;
     \\
     &\lim_{x\to0}U^{(1)}_- = 0; & &\lim_{x\to\infty} U^{(1)}_+ = u^*;
     \\
     &\lim_{x\to0}\left[U^{(s)}_-\mathcal{K}^{(s)}\right]_+ = \displaystyle -\sum_{j=1}^{J^{(s)}}  w^{(s)}_{j,-}b_{j,-}^{(s)} ; &  &\lim_{x\to\infty}\left[U^{(s)}_-\mathcal{K}^{(s)}\right]_+ = 0 ;
     \\
     &\lim_{x\to0}\left[U^{(s)}_+\mathcal{K}^{(s)}\right]_- = \displaystyle -\sum_{j=1}^{J^{(s)}}  w^{(s)}_{j,+}b_{j,+}^{(s)};&      &\lim_{x\to\infty}\left[U^{(s)}_+\mathcal{K}^{(s)}\right]_- = 0 ;
     \\
     &\lim_{x\to0}\left[U^{(s)}_-\mathcal{K}^{(s)}\right]_- = \displaystyle \sum_{j=1}^{J^{(s)}}  w^{(s)}_{j,-}b_{j,-}^{(s)}; &      &\lim_{x\to\infty}\left[U^{(s)}_+\mathcal{K}^{(s)}\right]_+ = 0 ;
     \\
    &\lim_{x\to0}\left[\mathcal{K}^{(1)}\right]_\pm = \displaystyle \mp\sum_{j=1}^{J^{(1)}}b_{j,-}^{(1)}; &    &\lim_{x\to\infty}\left[\mathcal{K}^{(1)}\right]_\pm = 0 ; 
     \\
     &\lim_{x\to0}\left[\Pi\mathcal{K}^{(1)}\right]_\pm = \displaystyle \mp\left[\mathcal{K}^{(1)}(x_p^-) + \sum_{j=1}^{J^{(1)}}\Pi\left(\lambda_{j,-}^{(1)}\right)b_{j,-}^{(1)}\right]; &    &\lim_{x\to\infty}\left[\Pi\mathcal{K}^{(1)}\right]_\pm = 0.
\end{align*}
}


\FloatBarrier

%% file: append_4x4.tex
\section{General case: dealing with the \texorpdfstring{$4\times 4$}{4×4} matrix Wiener--Hopf equation}
\label{sect_4x4_pole_removal}
In this section, we provide a sketch of the solution to the matrix Wiener--Hopf equation for the asymmetric waveguide \equ{eq_matrix_eq_44}. In this general case, the $4\times 4$ matrix leads to the following system of equations:
\begin{align}
    &-\U_{\ell_1-1}\Phi(x,-n_1-1)+\U_{\ell_1-2}\Phi(x,-n_1)=0 \label{eq_asym_syst1}
    \\
    &-\U_{\ell_2-1}\Phi(x,n_2+1)+\U_{\ell_2-2}\Phi(x,n_2)=0
    \\
    &-\U_{\ell_0-1}\Psi(x,-n_1)+\Psi(x,n_2)+\U_{\ell_0-2}\Psi(x,-n_1-1)=R_1(x)
    \\
    &-\U_{\ell_0-1}\Psi(x,n_2)+\Psi(x,-n_1)+\U_{\ell_0-2}\Psi(x,n_2+1)=R_2(x) \label{eq_asym_syst4}
\end{align}
where $R_1,\ R_2$ are defined in \equ{eq_R1_x4}-\equ{eq_R2_x4}. In a similar manner as in the symmetric case, we can solve each equation in \equ{eq_asym_syst1}-\equ{eq_asym_syst4} using the pole removal technique (the dependence of the functions on $x$ is omitted for brevity but implied):
\begin{align}
    U_+(-n_1-1) &=  U_+(-n_1)\mathcal{M}^{(1)}  
    +
    \sum_{j=1}^{M^{(1)}} a^{(1)}_j\mathcal{F}^{(1)}_j
    + \mathcal{G}^{(1)}_0 . \label{eq_asym_sol_pr1}
    \\
    U_+(n_2+1) &=  U_+(n_2)\mathcal{M}^{(2)}
    +
    \sum_{j=1}^{M^{(2)}} a^{(2)}_j\mathcal{F}^{(2)}_j
    + \mathcal{G}^{(2)}_0 . \label{eq_asym_sol_pr2}
    \\
    U_+(-n_1) &= -U_+(-n_1-1)\mathcal{M}^{(0)}-U_+(n_2)\mathcal{M}^{(3)} +  \sum_{j=1}^{M^{(0)}}  d^{(1)}_j\mathcal{F}^{(0)}_j
    +  \sum_{j=1}^{M^{(0)}}  c^{(2)}_j\mathcal{F}^{(3)}_j
    + \mathcal{G}_1^{(1)}, \label{eq_asym_sol_pr3}
    \\
    U_+(n_2) &= -U_+(n_2+1)\mathcal{M}^{(0)} - U_+(-n_1)\mathcal{M}^{(3)} +  \sum_{j=1}^{M^{(0)}}  d^{(2)}_j\mathcal{F}^{(0)}_j
    +  \sum_{j=1}^{M^{(0)}}  c^{(1)}_j\mathcal{F}^{(3)}_j
    + \mathcal{G}_1^{(2)}.  \label{eq_asym_sol_pr4}
\end{align} 
Solutions \equ{eq_asym_sol_pr1}-\equ{eq_asym_sol_pr4} depend on $2M^{(0)}+M^{(1)}+M^{(2)}$ unknown constants:
\begin{alignat}{2}
    &a^{(1)}_j=U_+\left(\lambda_{j,+}^{(1)},-n_1\right), &\quad &a^{(2)}_j=U_+\left(\lambda_{j,+}^{(2)},n_2\right),
    \\
    &d^{(1)}_j=U_+\left(\lambda_{j,+}^{(0)},-n_1-1\right), &\quad &d^{(2)}_j=U_+\left(\lambda_{j,+}^{(0)},n_2+1\right),
    \\
    &c^{(1)}_j=U_+\left(\lambda_{j,+}^{(0)},-n_1\right), &\quad & c^{(2)}_j=U_+\left(\lambda_{j,+}^{(0)},n_2\right);
\end{alignat}
and known functions
\begin{align}
\mathcal{M}^{(s)}(x) &= \frac{\U_{\ell_s-2}(z(x))}{\U_{\ell_s-1}(z(x))} = -x\frac{P^{(s)}(x)}{Q^{(s)}(x)}, \quad s\in\{1,2\}
\label{eq_4x4_Ms}
    \\
\mathcal{M}^{(0)}(x) &= -\frac{\U_{\ell_0-2}(z(x))}{\U_{\ell_0-1}(z(x))} = x\frac{P^{(0)}(x)}{Q^{(0)}(x)}, \label{eq_4x4_M0}
    \\
    \mathcal{M}^{(3)}(x) &= -\frac{1}{\U_{\ell_0-1}(z(x))} = \frac{(-1)^{\ell_0}x^{\ell_0-1}}{Q^{(0)}(x)},
    \label{eq_4x4_M3}
\\
    \mathcal{F}^{(s)}_j(x) &= b_{j,-}^{(s)}\frac{x}{x-\lambda_{j,-}^{(s)}} - b_{j,+}^{(s)}\frac{x}{x-\lambda_{j,+}^{(s)}}, \quad b_{j,\pm}^{(s)} = - \frac{P^{(s)}(\lambda_{j,\pm}^{(s)})}{Q_{j,\pm}^{(s)}(\lambda_{j,\pm}^{(s)})}, \quad s=\{1,2\}
    \label{eq_4x4_Fs1}
    \\
    \mathcal{F}^{(0)}_j(x) &= b_{j,-}^{(0)}\frac{\lambda_{j,-}^{(0)}}{x-\lambda_{j,-}^{(0)}} + b_{j,+}^{(0)}\frac{\lambda_{j,+}^{(0)}}{x-\lambda_{j,+}^{(0)}}, \quad 
    b_j^{(0)\pm} = \frac{P^{(0)}(\lambda_{j,\pm}^{(0)})}{Q_{j,\pm}^{(0)}(\lambda_{j,\pm}^{(0)})}, 
    \\
    \mathcal{F}^{(3)}_j(x) &= b_j^{(3)-}\frac{(\lambda_{j,-}^{(0)})^{\ell_0-1}}{x-\lambda_{j,-}^{(0)}} + b_{j,+}^{(3)}\frac{(\lambda_{j,+}^{(0)})^{\ell_0-1}}{x-\lambda_{j,+}^{(0)}}, \quad b_{j,\pm}^{(3)} =  \frac{(-1)^{\ell_0}}{Q_{j,\pm}^{(0)}(\lambda_{j,\pm}^{(0)})},
    \label{eq_4x4_Fs4}
\end{align}
where the rational functions $P^{(s)}(x),\ Q^{(s)}(x), \ s\in\{0,2\}$ in \equ{eq_4x4_Ms}-\equ{eq_4x4_M3} are defined similarly to \equ{eq_P}-\equ{eq_Q}, and the rational functions $Q_{j,\pm}^{(s)}(x),\ s\in\{0,2\}$ in \equ{eq_4x4_Fs1}-\equ{eq_4x4_Fs4} are defined similarly to \equ{eq_Q_j}. 
The known functions $\mathcal{G}_0^{(s)}(x)$ and $\mathcal{G}_1^{(s)}(x)$ in \equ{eq_asym_sol_pr1}-\equ{eq_asym_sol_pr4} are given by
\begin{align}
    \mathcal{G}_0^{(s)}(x) &= \operatorname{s}_p(\ell_s)\sum_{j=1}^{M^{(s)}}b_{j,-}^{(s)} \frac{x}{x-\lambda_{j,-}^{(s)}}, \quad  s=\{1,2\} \label{eq_4x4_G0}
    \\
    \mathcal{G}_1^{(1)}(x) &= -\operatorname{s}_p(\ell_2)\frac{x}{x-x_p^-} -
    \operatorname{s}_p(\ell_2) \left[\Pi\mathcal{M}^{(3)}\right]_+ - \operatorname{s}_p(\ell_1-1)
    \left[\Pi\mathcal{M}^{(0)}\right]_+,
    \\
    \mathcal{G}_1^{(2)}(x) &= -\operatorname{s}_p(\ell_1)\frac{x}{x-x_p^-} -
    \operatorname{s}_p(\ell_1) \left[\Pi\mathcal{M}^{(3)}\right]_+ - \operatorname{s}_p(\ell_2-1)
    \left[\Pi\mathcal{M}^{(0)}\right]_+,
\end{align}
where 
\begin{align}
    \left[\Pi\mathcal{M}^{(3)}\right]_+ &= \mathcal{M}^{(3)}(x_p^-)\frac{x_p^-}{x-x_p^-} +\sum_{j=1}^{M^{(0)}}\Pi (\lambda_{j,-}^{(0)}) b_j^{(3)-} \frac{(\lambda_{j,-}^{(0)})^{\ell_0-1}}{x-\lambda_{j,-}^{(0)}}, \label{eq_4x4_PiK3}
    \\
    \left[\Pi\mathcal{M}^{(0)}\right]_+ &= \mathcal{M}^{(0)}(x_p^-)\frac{x_p^-}{x-x_p^-} +\sum_{j=1}^{M^{(0)}}\Pi (\lambda_{j,-}^{(0)})b_{j,-}^{(0)} \frac{\lambda_{j,-}^{(0)}}{x-\lambda_{j,-}^{(0)}}.
    \label{eq_4x4_PiK0}
\end{align}
Note that in \equ{eq_asym_sol_pr1}-\equ{eq_asym_sol_pr4}, \equ{eq_4x4_G0} and \equ{eq_4x4_PiK3}-\equ{eq_4x4_PiK0} $M^{(0)}$ is the number of poles in \equ{eq_4x4_M0}, and $M^{(s)}$ is the number of poles in \equ{eq_4x4_Ms}.

Then we find
\begin{align}
    U_+(-n_1) &= -\frac{\mathcal{M}^{(3)}U_+(n_2) +\mathcal{Q}_1}{1+\mathcal{M}^{(0)}\mathcal{M}^{(1)}},
    \label{eq_u_pos_n1}
\\
   U_+(n_2) &= \frac{\mathcal{Q}_1 + (1+\mathcal{M}^{(0)}\mathcal{M}^{(1)})\mathcal{Q}_2}{\Theta}, 
   \label{eq_u_pos_n2}
\end{align}
where
\begin{equation}
    \frac{1}{\Theta} = \frac{1}{\left(1+\mathcal{M}^{(0)}\mathcal{M}^{(1)}\right)\left(1+\mathcal{M}^{(0)}\mathcal{M}^{(2)}\right) + \left(\mathcal{M}^{(3)}\right)^2}, \label{eq_denominator}
\end{equation}
and
\begin{align}
   \mathcal{Q}_1 &=  \mathcal{M}^{(0)}\sum_{j=1}^{M^{(1)}}  a^{(1)}_j\mathcal{F}^{(1)}_j  -  \sum_{j=1}^{M^{(0)}}  d^{(1)}_j\mathcal{F}^{(0)}_j
    -  \sum_{j=1}^{M^{(0)}}  c^{(2)}_j\mathcal{F}^{(3)}_j
    - \left[\mathcal{G}_1^{(1)}-\mathcal{M}^{(0)}\mathcal{G}_0^{(1)}\right], \label{eq_unknowns1}
    \\
    \mathcal{Q}_2 &=  -\mathcal{M}^{(0)}\sum_{j=1}^{M^{(2)}}  a^{(2)}_j\mathcal{F}^{(2)}_j  +  \sum_{j=1}^{M^{(0)}}  d^{(2)}_j\mathcal{F}^{(0)}_j
    +  \sum_{j=1}^{M^{(0)}}  c^{(1)}_j\mathcal{F}^{(3)}_j
    + \left[\mathcal{G}_1^{(2)}-\mathcal{M}^{(0)}\mathcal{G}_0^{(2)}\right]. \label{eq_unknowns2}
\end{align}
The solution for $U_+(-n_1-1)$ and $U_+(n_2+1)$ is given by \equ{eq_asym_sol_pr1}-\equ{eq_asym_sol_pr2} and \equ{eq_u_pos_n1}-\equ{eq_u_pos_n2}. 

In a similar manner to the symmetric case, \equ{eq_denominator} has $M$ poles $\nu_i^+$ in the $+$ region. Therefore, the expression in the numerator of \equ{eq_u_pos_n2} must vanish at these poles to preserve the analyticity of $U_+(n_2)$. In this case, the values of $\nu_i^+$ can only be determined numerically. There are three possible scenarios:
\begin{enumerate}
    \item $\ell_0 \neq \ell_1,\ \ell_0 \neq \ell_2,\ \ell_0 \neq \ell_1\pm1,\ \ell_0 \neq \ell_2\pm1$

    Then \equ{eq_denominator} has $M = 2M^{(0)} + M^{(1)} + M^{(2)}$ poles to determine the unknowns in \equ{eq_unknowns1}-\equ{eq_unknowns2}, in a similar way to the symmetric case \equ{eq_algebraic_system_coeff}. Note that the number of unknowns is also equal to $M$, since $d_j^{(1,2)}$ can be expressed in terms of $a_j^{(1,2)}$, and $c_j^{(1,2)}$ in terms of \equ{eq_asym_sol_pr1}--\equ{eq_asym_sol_pr2}.
    
    \item $\ell_0 = \ell_1\ \text{and/or } \ell_0 = \ell_2$

    Then the difference from the previous case is that $a_j^{(1)} = c_j^{(1)}$ and/or $a_j^{(2)} = c_j^{(2)}$. Therefore, the number of unknowns is still equal to the number of poles.

    \item $\ell_0 = \ell_1\pm1\ \text{and/or } \ell_0 = \ell_2\pm1$

Then the number of poles \( M \) is smaller than the number of unknowns. However, we can complete the algebraic system using a similar method as in the symmetric case, as discussed at the end of \sect{sect_lenar_system}.
\end{enumerate}

%% file: append_bae.tex
\section{Boundary algebraic equations system derivation}
\label{appx_bae}

To find a solution for the problem \equ{eq_Helmholtz_discrete_sc}--\equ{eq_BC_symmetric2} in terms of the boundary algebraic equations system, we define the Helmholtz operator as
\begin{equation}
    \mathcal{L}[f_{m,n}] = f_{m+1,n} + f_{m-1,n} + f_{m,n+1} + f_{m,n-1} + (\Omega^2-4)f_{m,n}, \quad \{m,n\}\in\mathbb{Z}\times\mathbb{Z}_N.
    \label{equ_L_operator}
\end{equation}
From the properties of the operator $\mathcal{L}$, for any two functions $f_{m,n}$ and $w_{m,n}$ that satisfy \equ{equ_L_operator}, it follows (as in \cite{Poblet-Puig2015SuppressionEquations}) that
\begin{equation}
    \sum_{m,n\in\mathbb{Z}\times\mathbb{Z}_N} \mathcal{L}[f_{m,n}]\,w_{m,n} 
    = \sum_{m,n\in\mathbb{Z}\times\mathbb{Z}_N} \mathcal{L}[w_{m,n}]\,f_{m,n}.
    \label{eq_sums}
\end{equation}

From the boundary conditions \equ{eq_BC_symmetric1}-\equ{eq_BC_symmetric2}, for the scattered field $u^\text{sc}_{m,n}$ we write:
\begin{equation}
    \mathcal{L}[u^\text{sc}_{m,n}] =
    \left\{
    \begin{array}{ll}
        0, & \{m,n\} \notin\Gamma''; \\
        2u_{1,n}^{\text{sc}} + (u_{-1,n}^{\text{in}}+u_{1,n}^{\text{in}}) + \delta_{n,n_2}(u^\text{sc}_{0,n+1}+u_{0,n+1}^{\text{in}}) + \delta_{n,-n_1}(u^\text{sc}_{0,n-1}+u_{0,n-1}^{\text{in}}), & \{m,n\} \in\Gamma'',
    \end{array} 
    \right.
    \label{eq_bc_helm_op}
\end{equation}
where $\Gamma''$ is defined in \equ{eq_gamma_dashdash}.
From \equ{eq_green_system_eq}, for the tailored lattice Green's function we write:
\begin{equation}
    \mathcal{L}[G_{m,n}^{m_0,n_0}] = \delta_{m,m_0}\delta_{n,n_0}.
    \label{eq_g_helm}
\end{equation}
Taking $G_{m,n}^{m_0,n_0}$ as $f_{m,n}$ and $u^\text{sc}_{m,n}$ as $w_{m,n}$ in \equ{eq_sums}, and using \equ{eq_bc_helm_op}-\equ{eq_g_helm}, we obtain the solution \equ{eq_solution_BAE} for the scattered field at the point $(m_0,n_0)$.